# Nonlinear dynamic pressure beneath waves in water of intermediate depth: Theory and experiment

A.V. Slunyaev[1,2], A.V. Kokorina[1], M. Klein[3]

[1] Institute of Applied Physics, 46 Ulyanova Street, Nizhny Novgorod 603950, Russia
[2] National Research University-Higher School of Economics, 25 Bol'shaya Pechorskaya Street, Nizhny Novgorod 603950, Russia
[3] Hamburg University of Technology, Offshore Dynamics Group (M-14), Am Schwarzenberg-Campus 1, 21073 Hamburg, Germany

**Abstract**
The data of simultaneous measurements of the surface displacement produced by propagating planar waves in experimental flume and of the dynamic pressure beneath the waves are compared with the theoretical predictions based on different approximations for modulated potential gravity waves. Regular and irregular wave sequences in intermediate depths are considered. The efficiency of different models for reconstruction the pressure field from the known surface displacement time series (the direct problem) is investigated. A new two-component theory for weakly modulated weakly nonlinear waves is proposed showing the best performance among all applied models. Peculiarities of the vertical mode of the pressure second harmonic are discussed.

## 1. Introduction

Reliable long-term recordings of sea waves are required for many purposes. The registration of the sea surface wave movement with the help of bottom pressure sensors is advantageous from the practical point of view, as these sensors are relatively cheap, easy to deploy and are much less affected by storms and occasions of vandalism. However, the problem of reconstruction of the surface shape from the single-point measurement of a bottom sensor (the inverse problem) is nontrivial due to various reasons. The most straightforward solution of the inverse problem using the hydrostatic approximation is applicable to a very limited range of situations. The linear theory for potential waves is most frequently used to relate the surface displacement and the bottom pressure (e.g. Tsai et al, 2005), though significant deviations from the measurements have been emphasized for a long time (e.g. Cavaleri et al, 1978). Nonlinear effects intensify in shallower water, where the bottom pressure gauges are used most frequently; this greatly complicates the solution of the inverse problem. Nonlinear reconstruction methods have been suggested instead of the spectral transfer function (see DiMarco et al, 2001; Bonneton & Lannes, 2017, Vasan et al, 2017; Mouragues et al, 2019; and references therein). The majority of the related studies are focusing on the conditions from shallow to intermediate depth water. New nonlinear procedures for the reconstruction of extreme periodic waves in the water of constant depth were suggested in Clamond & Constantin (2013), Clamond (2013) and Clamond & Henry (2020). Weakly nonlinear theory for the reconstruction of the free-surface displacement taking into account wave vorticity (e.g. due to shear currents), was developed in Henry & Thomas (2018).

At the same time, the direct problem of evaluation of the pressure field beneath nonlinear irregular waves possesses its own importance. The dynamic pressure variations caused by the sea surface waves may be transferred to the bottom. Imprints of storms in the open ocean have been found in seismic signals, and the effect of microseisms is considered as a promising tool for distant registration of storms and typhoons [Hasselmann, 1963; Landès et al, 2010; Ardhuin et al, 2012; Kovalev et al, 2015]. The microseismic effect is significant in the deep ocean [Obrebski et al, 2012; Ying et al, 2014], while the conditions favorable for the strong seismic signal are still not understood in full [Glubokovskikh et al, 2021]. A better understanding of the pressure fields produced by intense surface waves is necessary for more efficient diagnostics of sea conditions, as well as for the improvement of the methods to solve the inverse problem. Approximate analytic solutions for particular surface wave patterns (e.g. long-wave solitons in shallow water, weakly nonlinear cross seas, wave groups in relatively deep water, etc., see Pellet et al, 2017; Slunyaev et al, 2018; Touboul & Pelinovsky, 2018) are useful for better qualitative understanding, and may be used as benchmarks.

One may presume that arrays of pressure gauges distributed at different locations at a few depths might be a more efficient instrument for the reliable and inexpensive registration of surface waves. Then the interest to the pressure field beneath sea waves should not be limited only to the region near the bottom.

In this paper, we study theoretically and experimentally the relation between the water surface displacement and the pressure field beneath nonlinear waves. The conditions correspond to relatively deep water. Since the fluid velocities were not recorded in the experiments, the wave data is incomplete. We solve the direct problem and calculate the pressure fields within the linear and weakly nonlinear theories, using the time series of the surface displacement at a given location as the input data.

The paper is organized as follows. The experimental setup is described in Sec. 2. The governing equations and the reference (linear) theory are presented in Sec. 3 including comparison with the experimental data. The weakly nonlinear weakly dispersive theory for the pressure in the deep-water limit is presented and discussed in Sec. 4. Its efficiency is checked against the laboratory measurements at the end of the section. A new theory for the pressure field is presented in Sec. 5, discussing peculiarities of the solution for the pressure nonlinear harmonics. The new theory is verified against the experimental data in Sec. 6. A brief summary concludes the paper.

## 2. Experimental setup and the wave parameters

The experimental tests have been performed in the small wave tank at the Technical University of Berlin. The dimensions of the tank are: length 15 m, width 0.3 m, the bottom is flat, and the water depth is $h = 0.4$ m. In total six series were simulated, five with regular waves (A-E) and one with irregular waves (F), see Table 1. The irregular train was a tailored wave sequence which reproduced a rogue wave, recorded off Yura harbor in the Sea of Japan (Mori et al., 2002). The reconstruction of the Yura wave has been previously presented in Clauss et al. (2007). The waves were generated with a piston-type wave maker at one side of the tank; an absorbing beach was installed at the other side. The pressure sensor was located approximately in the middle of the wave flume. In each test run, the dynamic pressure was recorded at one given depth $d > 0$, simultaneously with the surface displacement exactly above the probe. Then the experiment was repeated for each new vertical position of the pressure sensor, $z = -d_j$, $j = 1,…, N$. The sampling frequency of the experimental measurements was set to 200 Hz being sufficient for high resolution of records. In the experiments A-E with regular waves, the pressure time series at each depth $d_j$ is analyzed with the corresponding simultaneously measured series of the surface displacement. Only one record of the surface displacement is used in the case F with modulated waves. For this study, only the time series from the pressure sensors are considered, which did not emerge from the water in the wave troughs. The static pressure has been automatically removed from the records, so that its zero value corresponds to the still water condition. In the subsequent processing, the mean values of the surface displacement and of the pressure are not subtracted from the data.

The experimental runs are characterized by different peak frequencies $T_p = 2\pi/\omega_p$, which provide deeper (A-B and F) or shallower conditions (C-E), characterized by the dimensionless depth $k_p h$. The peak wavenumber $k_p$ is calculated from the linear dispersion relation for gravity water wave, see Table 1. As aforementioned, the selected sampling frequency rate of the measurements was sufficient to measure at least several nonlinear superharmonics. For the series with regular waves A-E, the wave amplitude is characterized by the height $H$ divided by 2 (see Table 1). The root-mean-square surface displacement $\eta_{rms}$ is also given for all the series.

Several examples of the surface displacement time series (the upper blue curve) and the pressure records at a few horizons (the black curves below) are shown in Fig. 1 for the experiments with steeper waves B, E, F. For the analysis, only the leading parts of the regular wave sequences were used (see the red wave segments in Fig. 1a,b), which did not show signs of wave reflection from the absorbing beach. The sample lengths were selected manually to provide nearly periodic boundary conditions in order to ensure faster decay of Fourier transforms at high frequencies. A longer piece containing significant waves was examined in the case F. In this experiment a very high wave occurs at $t \approx 36$ s with the crest amplitude $A_{cr} \approx 8.8$ cm and the trough amplitude $A_{tr} \approx 4.2$ cm (see also Fig. 3a below). This wave is characterized by large steepness parameters, $k_p H/2 \approx 0.49$ and $k_p A_{cr} \approx 0.66$.

## 3. Basic equations and the linear approximation

In the present study, we employ the equations of hydrodynamics for potential flows in ideal fluid. The water bed corresponds to the horizon $z = -h$, the free water surface is described by the function $z = \eta(x,t)$ with the rest level $z = 0$, waves are planar and may propagate along the horizontal axis $Ox$. The velocity potential $\varphi(x,z,t)$ satisfies the Laplace equation in the volume $-h \leq z \leq \eta(x,t)$,

$$\frac{\partial^2 \varphi}{\partial x^2} + \frac{\partial^2 \varphi}{\partial z^2} = 0, \qquad -h \leq z \leq \eta. \tag{1}$$

The non-leaking bottom boundary condition applies,

$$\frac{\partial \varphi}{\partial z} = 0, \qquad z = -h. \tag{2}$$

Two surface boundary conditions relate the velocity potential and the surface displacement,

$$g\eta + \frac{\partial \varphi}{\partial t} = -\frac{1}{2}\left(\left(\frac{\partial \varphi}{\partial x}\right)^2 + \left(\frac{\partial \varphi}{\partial z}\right)^2\right) \quad \text{at} \quad z = \eta(x,t), \tag{3}$$

$$\frac{\partial \varphi}{\partial z} - \frac{\partial \eta}{\partial t} = \frac{\partial \eta}{\partial x}\frac{\partial \varphi}{\partial x} \quad \text{at} \quad z = \eta(x,t), \tag{4}$$

where $g$ is the acceleration due to gravity. The total pressure beneath the waves $P_{tot}$ is calculated according to the Bernoulli law as

$$\frac{1}{\rho}P_{tot} = -\frac{\partial \varphi}{\partial t} - \frac{1}{2}\left(\left(\frac{\partial \varphi}{\partial x}\right)^2 + \left(\frac{\partial \varphi}{\partial z}\right)^2\right) - gz, \tag{5}$$

where $\rho$ is the water density. The conditions (3) and (5) imply that the total pressure is zero on the surface, $P_{tot}(x,z=\eta,t) \equiv 0$. The quantity $p$

$$p = \frac{1}{\rho}P_{tot} + gz = -\frac{\partial \varphi}{\partial t} - \frac{1}{2}\left(\left(\frac{\partial \varphi}{\partial x}\right)^2 + \left(\frac{\partial \varphi}{\partial z}\right)^2\right) \tag{6}$$

is the normalized dynamic pressure, which characterizes the excess of the total pressure over the hydrostatic pressure. In what follows the "pressure" will refer to the quantity $p$.

The solution of the linearized equations of hydrodynamics may be straightforwardly obtained with the help of the Fourier transform (e.g. Dean & Dalrymple, 2010). For one progressive wave with the amplitude $A_0$, the wavenumber $k_0$ and the cyclic frequency $\omega_0$, the solution has the form

$$\eta(x,t) = A_0 \cos(\omega_0 t - k_0 x), \tag{7}$$

$$\varphi(x,z,t) = -\frac{g}{\omega_0} A_0 \sin(\omega_0 t - k_0 x) \frac{\cosh(k_0(z+h))}{\cosh(k_0 h)}, \tag{8}$$

$$p(x,z,t) = gA_0 \cos(\omega_0 t - k_0 x) \frac{\cosh(k_0(z+h))}{\cosh(k_0 h)}, \tag{9}$$

where $\omega_0$ and $k_0$ are constrained by the dispersion relation $\omega_0 = \omega(k_0)$,

$$\omega(k) = \sqrt{gk\sigma}, \quad \sigma = \tanh kh. \tag{10}$$

If many waves compose the wave field, then the linear solution is represented by a linear superposition of (7)-(10) for waves with arbitrary phases. For the given time series of the surface displacement $\eta(t)$ of unidirectional waves, the linear dynamic pressure solution reads

$$p(x,z,t) = \frac{g}{2\pi} \int_{-\infty}^{\infty}\int_{-\infty}^{\infty} \eta(\tau) e^{i\omega(t-\tau)-ik(\omega)x} \frac{\cosh(k(\omega)(z+h))}{\cosh(k(\omega)h)} d\omega d\tau, \tag{11}$$

where the dependence $k(\omega)$ is prescribed by (10). The solution (11) may be straightforwardly transformed to the solution in the spatial domain. The linear dynamic pressure beneath progressive waves (11) attenuates with depth, so that at large depths the total pressure tends to hydrostatic.

In Fig. 2, the dynamic pressure calculated from the surface displacement time series $\eta(t)$ according to the linear solution (11) (black asterisks) is compared with the direct measurements of the pressure. The root-mean-square difference between the calculated, $p_{calc}$, and the experimentally measured, $p_{exp}$, pressures is used as an estimator of the accuracy of the theoretical description,

$$Err = \sqrt{\frac{\int (p_{calc} - p_{exp})^2 dt}{\int p_{exp}^2 dt}}, \tag{12}$$

where the integration by time is performed for the extracted time-series segments (shown with red curves in Fig. 1).

Unreasonably large errors being unevenly distributed with depth are observed in the small-amplitude case A (*Err* is above 10% for the deepest probe). Steeper waves over the same depth (case B) exhibit more regular dependence of *Err* with the error below 4% at the deepest points. The series C for shallower condition and about twice steeper waves than in the case A, is characterized by smaller values of the error as well. Therefore, we assume that large errors in the case A are most likely the result of the pressure gauge sensitivity limit. According to the linear solution (11), the errors (12) in the experiments B, D-F with intense waves decrease with depth. In the series D-F the function *Err* reaches about 20% close to the water surface. The best accuracy within 3% is observed in the case C (small amplitude waves in the shallower water). In the situation of steep irregular waves (Fig. 2F) the linear theory results in the error of about 17% right under the waves and gradually goes down to 10% at $k_p d \sim 1.4$.

Nonlinear theories for the pressure field are considered in the following. The effect of nonlinearity appears at different steps of calculating the dynamic pressure, as outlined below.

i) The surface displacement $\eta(t)$ should be transformed to the velocity potential at the water rest level, $\varphi(t,z=0)$ according to the nonlinear surface conditions (3) and (4).

ii) The solution of the Laplace equation is linear, it yields the potential at the locations of the pressure sensors, $\varphi(t,z=-d)$.

iii) The sought dynamic pressure is given by the nonlinear Bernoulli law (6).

Since the water is deep or moderately deep in this test campaign ($k_p h > 1.7$), the weakly nonlinear Stokes-wave theory may be applied, when the solution is decomposed into the asymptotic series describing the dominant and multiple nonlinear wave harmonics. In addition, waves will be assumed slowly modulated. These approximations often lead to a sufficiently accurate description of realistic sea waves (e.g. Shemer et al, 2010, Xiao et al, 2013; Adcock & Taylor, 2016).

## 4. The framework of the deep-water Dysthe theory for weakly nonlinear weakly modulated waves

The wave nonlinearity may be a reason of the poor accuracy, particularly at the horizons close to the surface. In this section, we apply the popular nonlinear Dysthe theory, which is valid for sufficiently deep basins, to describe the pressure field beneath surface waves.

The Dysthe theory [Dysthe, 1979, Trulsen & Dysthe, 1996] is a nonlocal modified nonlinear Schrödinger equation for slowly modulated weakly nonlinear waves, which possesses an improved description of the induced nonlinear flow. According to the Laplace equation, longer perturbations of the velocity potential penetrate deeper. Therefore, the description of the pressure, caused by the large-scale induced flow, is important at deep horizons. The Dysthe model is popular in the hydrodynamic and oceanographic studies, as it turned out to be reasonably accurate in a number of tests (e.g. Shemer et al, 2010, Xiao et al, 2013; Adcock & Taylor, 2016; Tang et al, 2020); it requires relatively deep water basins. Importantly, two versions of the model which govern the wave evolution in time and the evolution in space are available (Trulsen, 2006).

The Dysthe theory is based on the assumptions of weak nonlinearity, $k_0 a \ll 1$, and weak dispersion, $\Delta k/k_0 \ll 1$, where $a$, $k_0$ and $\Delta k$ are the characteristic wave amplitude, the carrier wavenumber and the spectral width, correspondingly. The small quantities are of a similar order of magnitudes, $k_0 a \sim \Delta k/k_0 = O(\varepsilon)$, $\varepsilon \ll 1$. Hereafter $\varepsilon$ is used as a marker. The Dysthe theory implies that the characteristic modulation length is shorter than the water depth, what leads to the deep water condition, $k_0 h \geq \varepsilon^{-1}$. For unidirectional waves, the Dysthe theory describing the evolution of the complex amplitude $A(x,t)$ in space (the spatial form of the Dysthe equation) reads

$$i\left(\frac{\partial A}{\partial x} + \frac{1}{C}\frac{\partial A}{\partial t}\right) + \frac{k_0}{\omega_0^2}\frac{\partial^2 A}{\partial t^2} + k_0^3|A|^2 A - 8i\frac{k_0^3}{\omega_0}|A|^2\frac{\partial A}{\partial t} - 2i\frac{k_0^3}{\omega_0}A^2\frac{\partial A^*}{\partial t} - \frac{4k_0^3}{\omega_0^2}A\frac{\partial \overline{\varphi}}{\partial t}\bigg|_{z=0} = 0, \quad (13)$$

$$\frac{1}{C^2}\frac{\partial^2 \overline{\varphi}}{\partial t^2} + \frac{\partial^2 \overline{\varphi}}{\partial z^2} = 0, \qquad z \leq 0, \quad (14)$$

$$\frac{\partial \overline{\varphi}}{\partial z} = -\frac{\omega_0}{2C}\frac{\partial}{\partial t}|A|^2, \qquad z = 0, \quad (15)$$

$$\frac{\partial \overline{\varphi}}{\partial z} = 0, \qquad z = -h, \quad (16)$$

see e.g. Trulsen (2006) or Slunyaev et al (2014). Here, the wavenumber $k_0$ and the cyclic frequency $\omega_0$ of the carrier are related by the dispersion relation for deep water waves

$$\omega_0 = \sqrt{gk_0}, \tag{17}$$

instead of (10), and $C$ is the deep-water group velocity, $C = \omega_0/2/k_0$. The equation (13) describes the linear dispersion (17) exactly. The system (13)-(16) depends on the depth $h$ only through the bottom condition (16) for the long-scale component. The nonlinear and nonlinear-dispersive parts of (13) are accurate up to the order $O(\varepsilon^4)$. The induced flow potential $\bar{\varphi}(x,z,t) \sim O(\varepsilon^2)$ is one order of magnitude smaller than the velocity potential of the dominant wave. The fields $A$ and $\bar{\varphi}$ are slow functions of their coordinates $x$, $z$, $t$. The equation (14) is approximate; it is obtained from the Laplace equation (1), assuming that the induced flow propagates with the same velocity $C$ as the dominant wave modulation,

$$\frac{\partial \bar{\varphi}}{\partial x} = -\frac{1}{C}\frac{\partial \bar{\varphi}}{\partial t} + O(\varepsilon^4). \tag{18}$$

The condition (15) is accurate up to the order $O(\varepsilon^3)$, a similar change from the spatial to temporal derivative (18) was used to obtain it.

The physical fields are calculated from the solution of (13)-(16) according to the following reconstruction formulas (Trulsen, 2006; Slunyaev et al, 2014):

$$\eta(x,t) = \bar{\eta} + \eta^{(I)} + \eta^{(II)} + \eta^{(III)} + O(\varepsilon^4), \tag{19}$$

$$\eta^{(I)} = \mathrm{Re}(AE), \tag{20}$$

$$\eta^{(II)} = \frac{k_0}{2}\mathrm{Re}(A^2 E^2) + \frac{1}{2C}\mathrm{Im}\left(A\frac{\partial A}{\partial t}E^2\right), \tag{21}$$

$$\eta^{(III)} = \frac{3k_0^2}{8}\mathrm{Re}(A^3 E^3), \tag{22}$$

$$\bar{\eta} = -\frac{1}{2\omega_0 C}\left.\frac{\partial \bar{\varphi}}{\partial t}\right|_{z=0}, \tag{23}$$

$$\varphi(x,z,t) = \bar{\varphi} + \varphi^{(I)} + \varphi^{(II)} + \varphi^{(III)}, \tag{24}$$

$$\varphi^{(I)} = -\frac{\omega_0}{k_0}\mathrm{Im}(AE)e^{k_0 z} - \frac{1}{k_0}\mathrm{Re}\left[\frac{\partial A}{\partial t}E\right](1 - 2k_0 z)e^{k_0 z} \tag{25}$$

$$-\omega_0 k_0 \mathrm{Im}\left[A|A|^2 E\right]\left(\frac{3}{8} - k_0 z\right)e^{k_0 z} + \frac{1}{\omega_0 k_0}\mathrm{Im}\left[\frac{\partial^2 A}{\partial t^2}E\right](1 - k_0 z + 2k_0^2 z^2)e^{k_0 z},$$

$$\varphi^{(II)} = 0, \tag{26}$$

$$\varphi^{(III)} = 0, \tag{27}$$

$$E \equiv \exp(i\omega_0 t - i k_0 x). \tag{28}$$

The upper indices from (I) to (III) in (19) and (24) numerate the Fourier harmonics; the solution takes into account the terms of the surface displacement and of the fluid velocity $\nabla \varphi$ up to the order $O(\varepsilon^3)$.

The asymptotic solution for the dynamic pressure may be calculated according to (6) as follows,

$$p = \bar{p} + p^{(I)} + p^{(II)} + p^{(III)} + O(\varepsilon^4), \tag{29}$$

$$p^{(I)} = \frac{\omega_0^2}{k_0}\mathrm{Re}(AE)e^{k_0 z} + 2\omega_0 \mathrm{Im}\left(\frac{\partial A}{\partial t}E\right)z e^{k_0 z} \tag{30}$$

$$+\left[\omega_0^2 k_0 \mathrm{Re}\left(A|A|^2 E\right)\left(\frac{3}{8} - k_0 z\right)e^{k_0 z} - \mathrm{Re}\left(\frac{\partial^2 A}{\partial t^2}E\right)z(1 + 2k_0 z)e^{k_0 z}\right],$$

$$p^{(II)} = 0, \qquad (31)$$

$$p^{(III)} = 0. \qquad (32)$$

$$\bar{p} = -\frac{1}{2}\omega_0^2 |A|^2 e^{2k_0 z} - \frac{\partial \bar{\varphi}}{\partial t} - \omega_0 \operatorname{Im}\left[A^* \frac{\partial A}{\partial t}\right](1 + 2k_0 z)e^{2k_0 z}. \qquad (33)$$

Note that within the standard Dysthe theory (accurate to $O(\varepsilon^3)$), the terms of the second and third nonlinear harmonics do not appear in the expressions for the velocity potential and the pressure. The difference-frequency harmonic for the pressure, $\bar{p}$ (33), consists of one term of the order $O(\varepsilon^2)$ (the first summand) and two terms of the order $O(\varepsilon^3)$, which vanish if the wave is uniform. The second term in (33) is assumed to slowly attenuate with depth with characteristic scale $\sim \varepsilon^{-1} k_0^{-1}$, while the two remaining terms in $\bar{p}$ decay with depth quickly, with the scale $O(1)$.

With the purpose to calculate the pressure (29) for the given surface displacement $\eta(t)$, the complex amplitude $A(t)$ should be computed first. This may be done with the help of an iterative procedure which inverts (19) by optimizing the carrier frequency $\omega_0$ and the fitting function $A(t)$ to achieve the best fit for the surface displacement. In our computations, the Fourier transform of the amplitude function $A(t)$ was allowed to have nonzero values within the interval from $\omega_p/2$ to $3\omega_p/2$, assuming that the free wave component is confined within this frequency band. The pressure term in (33), which is produced by the induced flow $\bar{\varphi}$, is calculated using the solution of (14)-(16) in the form

$$\frac{\partial \bar{\varphi}}{\partial t} = \frac{\omega_0}{4\pi} \int_{-\infty}^{\infty}\int_{-\infty}^{\infty} \omega \coth\left(\omega \frac{2\omega_0 h}{g}\right) |A(\tau)|^2 \frac{\cosh\left(\omega \frac{2\omega_0(z+h)}{g}\right)}{\cosh\left(\omega \frac{2\omega_0 h}{g}\right)} e^{i\omega(t-\tau)} d\omega d\tau. \qquad (34)$$

The solution (34) contains the direct and inverse Fourier transforms; $\omega$ is the frequency in the Fourier domain. In the limit $k_0 h \to \infty$ the function *coth* in (34) is similar to the Heaviside step function, but with a discontinuity in the point $\omega = 0$; hence the mean of the function $|A(t)|^2$ should be zero to avoid secular solutions.

The full circles in Fig. 2 represent the comparison between the pressure fields calculated from the surface displacement using (29)-(33), and directly measured in the experiments. The corresponding points in deep-water cases A and B follow rather unevenly and do not show a definite advantage of the Dysthe model over the linear solution. Under the shallower conditions and large wave nonlinearity (cases D, E), the deep-water theory (29)-(33) describes the actual dynamic pressure in the upper layer $|k_p z| < 0.5$ with slightly better accuracy than the linear theory, while deeper the error seems to grow exponentially and reaches unacceptable magnitudes of the order of 30%. In the case C with smaller waves, the Dysthe theory at shallow horizons does not show a better result than the linear solution, but yields much worse description at deeper horizons.

In the experiment with irregular waves (series F), the error within the Dysthe theory starts from about 30% close to the surface and further grows to about 50% at $k_p z = -1.4$. In Fig. 3a, the surface displacement, reconstructed from the complex amplitude $A(t)$ according to (19)-(23) (red line), is compared with the original record (green line). Two corresponding pressure records are compared with the solution (29)-(33) in Figs. 3b,c. In Fig. 3a one may observe that the theory not only underestimates the sharp crests of steep waves (what could be expected due to the truncation of the series for the wave harmonics), but does not reproduce some significant wave patterns (see the instant $t \approx 30$ s). It turns out that the long-scale elevation component $\bar{\eta}$ which follows from the deep-water Dysthe theory is drastically smaller than the experimental value. The most significant differences in Figs. 3b,c seem to correspond to the locations of poor reconstruction of the water surface which may be found in Fig. 3a. It is interesting to note that the linear solution (blue curves) seems to follow the experimental data in Fig. 3b,c even better, though it does not reproduce the shapes of large oscillations.

We can suggest a few ideas why the Dysthe theory fails to describe well the experimental pressure fields. First, it takes into account the finiteness of depth only through the induced flow (14)-(16) due to the bottom condition (16). Other contributors to $p$ in (29) do not depend on $h$ at all, what may be particularly critical in the shallower cases C–E. Another related reason may be the total absence of nonlinear harmonics in the pressure solution (see (31), (32)). The second harmonic was clearly observed in the experimental pressure records (will be discussed further in Sec. 6). The unaccounted long-scale movement of water in the experimental basin may be a next source of discrepancy between the theory and the measurements.

The property of abnormally weak nonlinearity of the velocity potential of regular deep-water Stokes waves is well-known, see e.g. Fenton (1985). The second-harmonic term appears in the order $O(\varepsilon^4)$, and the term of the triple frequency appears in the order $O(\varepsilon^5)$:

$$\sqrt{\frac{k_0^3}{g}}\varphi_{Stokes} = \mu\sin\theta e^{k_0 z} - \mu^3\frac{1}{8}\sin\theta e^{k_0 z} + \mu^4\frac{1}{2}\sin 2\theta e^{2k_0 z} \qquad (35)$$

$$+ \mu^5\left(-\frac{7}{12}\sin\theta e^{k_0 z} + \frac{1}{12}\sin 3\theta e^{3k_0 z}\right) + O(\varepsilon^6), \quad \theta \equiv \omega t - k_0 x,$$

$$\frac{\omega}{\sqrt{gk_0}} = \left(1 + \frac{1}{2}\mu^2 + \frac{1}{2}\mu^4\right).$$

Here $k_0$ is the wavenumber, $\omega$ is the observed (nonlinear) frequency, $\mu$ is the alternative steepness parameter $\mu \equiv k_0 A_0 \sim O(\varepsilon)$, where $A_0$ is the amplitude of the first harmonic. (The asymptotic change $\mu \to \mu - 3/8\,\mu^3 - 422/384\mu^5 + O(\varepsilon^7)$ should be performed to obtain the relation in the form presented in Fenton (1985).) The coefficient at the term of the order $O(\mu^3)$ in (35) agrees with (25), when the difference between $\omega$ and $\omega_0$ is taken into account (see Eq. (10) in [Slunyaev et al, 2014]). When the deep-water solution (35) is substituted in (6), the nonlinear part of the Bernoulli formula does not yield high-frequency nonlinear harmonics in the low asymptotic orders:

$$\frac{1}{2}\left(\left(\frac{\partial\varphi}{\partial x}\right)^2 + \left(\frac{\partial\varphi}{\partial z}\right)^2\right) = \frac{g}{k_0}\left[\frac{1}{2}\mu^2\left(1 - \frac{1}{4}\mu^2\right)e^{2k_0 z} + \mu^5 e^{3k_0 z}\cos\theta\right] + O(\varepsilon^6). \qquad (36)$$

The Dysthe theory (29)-(33) is consistent with (36) up to the order $O(\varepsilon^4)$ when the wave is uniform ($\partial A/\partial t = 0$).

In the finite-depth framework, the Stokes wave's potential owns multiple nonlinear harmonics, which appear in the corresponding asymptotic orders of the small steepness (i.e., the second harmonic is present in the order $O(\varepsilon^2)$, etc.). The second harmonic of the dynamic pressure is significant in our experimental measurements of steep waves. Hence, a more accurate theory is required, which would take into account three effects: the wave nonlinearity, the wave dispersion, and the finite depth. In the next section, we develop an asymptotic nonlinear dispersive theory taking into account the finite depth, and treating the induced nonlinear flow in the spirit of Dysthe's approach [Dysthe 1979].

## 5. Derivation of the weakly nonlinear theory for modulated waves in a finite-depth basin in the presence of a long wave

In this section, we derive a weakly nonlinear theory for the pressure beneath surface gravity waves which propagate in a basin of relatively large but finite depth; a smaller-amplitude long wave may also be present in the basin (a 2-component theory, 2CT). This problem formulation arises from the typical situation in an experimental flume, where parasitic long waves may be excited by the wave maker or caused by incomplete wave absorbing at the other side of the tank.

The governing system of equations is given by (1)-(4). The leading order solution is supposed having the form of a combination of a wave train (subscript '*train*') and a long wave (subscript '*long*'). In line with the previous sections, the first wave is assumed small in amplitude having the carrier wavenumber $k_0$ and the cyclic frequency $\omega_0$, $k_0\eta_{train} \sim O(\varepsilon)$, $k_0\omega_0/g\, \varphi_{train} \sim O(\varepsilon)$, $\varepsilon \ll 1$. The wave modulations are assumed slow, with the horizontal scale $L_{train}$ much longer than the wave length $2\pi/k_0$, $k_0 L_{train} \sim O(\varepsilon^{-1})$. Such relation between nonlinear and dispersive effects is typical for the NLS theory (e.g. Slunyaev, 2005; Trulsen, 2006) and was assumed in Sec. 4. The corresponding characteristic time scale of the modulation is denoted as $T_{train}$, and its vertical scale is $D_{train}$. Then $k_0 D_{train} \sim k_0 L_{train} \sim O(\varepsilon^{-1})$ due to the Laplace equation, and $\omega_0 T_{train} \sim k_0 L_{train} \sim O(\varepsilon^{-1})$, assuming that the modulation propagates with the group velocity which is of the order of 1.

Let us estimate the scales of the long-wave component which may be excited in the laboratory flume by the primary waves. In this paragraph, we use dimensionless quantities for brevity, where the time is scaled using $\omega_0$, and the surface displacement and coordinates are scaled with the use of $k_0$. According to the linearized surface boundary conditions (the terms to the left-hand-side of (3)-(4)), the typical scales of the long wave are related as follows, $D_{long} \sim T_{long}^2$ and $\varphi_{long} \sim T_{long}\eta_{long}$; due to the Laplace equation $D_{long} \sim L_{long}$. If one assumes that the long wave is generated by the second-order induced fluid motions, then either $\varphi_{long} \sim O(\varepsilon^2)$ or $\eta_{long} \sim O(\varepsilon^2)$ depending on the particular mechanism of excitation. Due to the relation $\varphi_{long} \sim T_{long}\eta_{long}$ ($T_{long}$ is assumed to be a large scale), the biggest magnitudes of $\varphi_{long}$ and $\eta_{long}$ are provided when $\eta_{long} \sim O(\varepsilon^2)$; this particular estimation is assumed hereafter. Under unsteady but homogeneous conditions, the horizontal scales of the long wave and of the train modulation are similar, $L_{long} \sim L_{train} \sim O(\varepsilon^{-1})$, and then $D_{long} \sim O(\varepsilon^{-1})$, $T_{long} \sim O(\varepsilon^{-1/2})$, $\varphi_{long} \sim O(\varepsilon^{3/2})$. Consequently, the estimates follow: $\partial\varphi_{long}/\partial t \sim O(\varepsilon^2)$, $\partial\varphi_{long}/\partial x \sim O(\varepsilon^{5/2})$, $\partial\varphi_{long}/\partial z \sim O(\varepsilon^{5/2})$. If, alternatively, one assumes that the long wave is generated at the basin walls, then the time scales of the train and of the long wave agree, $T_{long} \sim T_{train} \sim O(\varepsilon^{-1})$; and therefore $D_{long} \sim O(\varepsilon^{-2})$, $L_{long} \sim O(\varepsilon^{-2})$, $\varphi_{long} \sim O(\varepsilon)$. Then the terms in (6) are estimated as follows, $\partial\varphi_{long}/\partial t \sim O(\varepsilon^2)$, $\partial\varphi_{long}/\partial x \sim O(\varepsilon^3)$, $\partial\varphi_{long}/\partial z \sim O(\varepsilon^3)$. In the both considered cases $\partial\varphi_{train}/\partial x \sim \partial\varphi_{train}/\partial z \sim O(\varepsilon)$, therefore the largest correction which may come from the nonlinear part of the Bernoulli formula (6) due to the nonlinear interactions between the long wave and the wave train is of the order $O(\varepsilon^{7/2}) < O(\varepsilon^3)$; it corresponds to the frequency band of the first harmonic.

Hereafter, we assume the long wave component to be sufficiently small in amplitude, so that it contributes to the solution linearly. The nonlinear solution for the weakly modulated wave train is found using the standard asymptotic theory (see e.g. Slunyaev, 2005), where the induced nonlinear wave flow is considered independently of the modulated wave, which is in line with the approach of Dysthe. Such a theory was derived by Brinch-Nielsen & Jonsson (1986) taking into account small terms up to the fourth order. We re-derive a lower-order version of that theory and use the result to calculate the pressure field. Details of the derivation are given in the Appendix.

The leading-order components of the surface displacement and of the velocity potential of the modulated train are

$$\eta_{train}(x,t) = \varepsilon \operatorname{Re}[A(x_1,t_1)E], \tag{37}$$

$$\varphi_{train}(x,t,z) = \varepsilon \operatorname{Re}[B(x_1,t_1)\Phi_{10}(z_0)E], \tag{38}$$

$$E(x_0,t_0) \equiv \exp(i\omega_0 t_0 - ik_0 x_0).$$

Here multiple coordinates and time are introduced with the purpose to decouple fast oscillations and the slow evolution,

$$\frac{\partial}{\partial x} = \frac{\partial}{\partial x_0} + \varepsilon \frac{\partial}{\partial x_1}, \quad \frac{\partial}{\partial z} = \frac{\partial}{\partial z_0} + \varepsilon \frac{\partial}{\partial z_1}, \quad \frac{\partial}{\partial t} = \frac{\partial}{\partial t_0} + \varepsilon \frac{\partial}{\partial t_1}. \tag{39}$$

The Laplace equation with the nonleaking condition at the bottom $z = -h$ yield the vertical mode structure $\Phi_{10}$, similar to the one of a plane wave (8) (see (A.5)). The surface displacement and the velocity potential are sought in the form of asymptotic series using the expansions in small parameter $\varepsilon$ for different wave harmonics characterized by the fast-oscillating function $E(x_0,t_0)$. We write the solution in the form of superposition of the wave harmonics as follows,

$$\eta = \eta^{(0)} + \eta^{(I)} + \eta^{(II)}, \quad \varphi = \varphi^{(0)} + \varphi^{(I)} + \varphi^{(II)}. \tag{40}$$

Only three harmonics will be considered, labeled with the corresponding upper indices. The wave train belongs to the band of the first harmonic, while the long wave corresponds to the low frequency band of the difference (or zeroth) harmonic.

The asymptotic expansions for $\eta$ and $\varphi$ are inserted into the governing equations (1)-(4). The boundary conditions at $z = \eta$ are approximated using the Taylor expansions in the vicinity of the water rest level $z = 0$. Finally, the terms which correspond to different harmonics (powers of $E$) and different orders of nonlinearity (powers of $\varepsilon$) are collected and considered separately, e.g. [Slunyaev, 2005]. Details of the asymptotic derivation are given in the Appendix. The eventual solution reads

$$\eta^{(I)} = \operatorname{Re}(AE), \tag{41}$$

$$\eta^{(II)} = \frac{k_0(3-\sigma^2)}{4\sigma^3} \operatorname{Re}(A^2 E^2), \tag{42}$$

$$\eta^{(0)} = \eta_{long} + \overline{\eta}, \quad \overline{\eta} = -\frac{k_0(1-\sigma^2)}{4\sigma}|A|^2 - \frac{1}{g}\frac{\partial \overline{\varphi}}{\partial t}\bigg|_{z=0} \tag{43}$$

$$+ \frac{(1-\sigma^2)(k_0 h(1+\sigma^2) - \sigma)}{4\sigma^2 V} \operatorname{Im}\left[A^* \frac{\partial A}{\partial t}\right],$$

$$\varphi^{(I)} = -\frac{g}{\omega_0} \operatorname{Im}(AE) \frac{\cosh(k_0(z+h))}{\cosh(k_0 h)} \tag{44}$$

$$- \frac{\omega_0}{2\sigma^2 k_0^2 V} \operatorname{Re}\left(\frac{\partial A}{\partial t} E\right)\left[(\sigma + k_0 h(1+\sigma^2))\frac{\cosh(k_0(z+h))}{\cosh(k_0 h)} - 2\sigma k_0(h+z_0)\frac{\sinh(k_0(z+h))}{\cosh(k_0 h)}\right],$$

$$\varphi^{(II)} = -\frac{3\omega_0(1-\sigma^4)}{8\sigma^4} \operatorname{Im}(A^2 E^2) \frac{\cosh(2k_0(z+h))}{\cosh(2k_0 h)}, \tag{45}$$

$$\varphi^{(0)} = \varphi_{long} + \overline{\varphi}, \tag{46}$$

$$V = \frac{g}{2\omega_0}(\sigma + k_0 h(1-\sigma^2)). \tag{47}$$

The term $V$ (47) is the group velocity of the wave train which may be calculated from (10) as $d\omega/dk$ for $k = k_0$; it tends to $C$, which was introduced in the previous section, in the limit of infinite depth, $V \to C = \omega_0/k_0/2$ when $k_0 h \to \infty$. The solution is written in the time domain, which is convenient for the comparison with single-point measurements.

The induced nonlinear surface displacement $\bar{\eta}$ (43) is always negative (wave set-down). The problem on the induced flow $\bar{\varphi}$ generalizes the problem (14)-(16); it consists of the following set of equations,

$$\frac{1}{V^2}\frac{\partial^2 \bar{\varphi}}{\partial t^2} + \frac{\partial^2 \bar{\varphi}}{\partial z^2} = 0, \qquad z \leq 0, \tag{48}$$

$$\frac{\partial \bar{\varphi}}{\partial z} = -\left(\frac{\omega_0}{2\sigma V} + \frac{k_0(1-\sigma^2)}{4\sigma}\right)\frac{\partial}{\partial t}|A|^2, \qquad z = 0, \tag{49}$$

$$\frac{\partial \bar{\varphi}}{\partial z} = 0, \qquad z = -h. \tag{50}$$

As discussed above, the long wave components $\eta_{long}$ and $\varphi_{long}$ enter the solution (43), (46) in the form of linear additives. The variables $\eta_{long}$ and $\varphi_{long}$, are constrained in this approximation according to the linear relation, and the corresponding solution may be written in the general form using the Fourier representation,

$$\eta_{long}(x,t) = \frac{1}{\sqrt{2\pi}}\int_0^\infty [a_{long}(\omega)\exp(i\omega t - ik(\omega)x) + c.c.]d\omega,$$

$$\varphi_{long}(x,t,z) = \frac{1}{\sqrt{2\pi}}\int_0^\infty \frac{g}{\omega}[ia_{long}(\omega)\exp(i\omega t - ik(\omega)x) + c.c.]\frac{\cosh(k(\omega)(z+h))}{\cosh(k(\omega)h)}d\omega,$$

$$a_{long}(\omega) = \frac{1}{\sqrt{2\pi}}\int_{-\infty}^\infty \eta_{long}(0,t)\exp(-i\omega t)dt. \tag{51}$$

Here, the dispersion relation $k(\omega)$ is given by (10), and $a_{long}(\omega)$ are the spectral amplitudes of the long wave component. If the long wave is characterized by a single frequency, the solution is given by (7)-(8).

The dynamic pressure beneath the waves is found according to the Bernoulli law (6). The solution may be written in the form of a superposition of three harmonics:

$$p = p^{(0)} + p^{(I)} + p^{(II)}, \tag{52}$$

$$p^{(I)} = g\,\mathrm{Re}(AE)\frac{\cosh(k_0(z+h))}{\cosh(k_0 h)} \tag{53}$$

$$+ \frac{g}{V}\mathrm{Im}\left(\frac{\partial A}{\partial t}E\right)\left[\frac{(z+h)\sinh(k_0(z+h))}{\cosh(k_0 h)} - \frac{h\sigma\cosh(k_0(z+h))}{\cosh(k_0 h)}\right],$$

$$p^{(II)} = \frac{gk_0(1-\sigma^2)}{4\sigma}\left[\frac{3(1+\sigma^2)}{\sigma^2}\frac{\cosh(2k_0(z+h))}{\cosh(2k_0 h)} - 1\right]\mathrm{Re}(A^2 E^2), \tag{54}$$

$$p^{(0)} = p_{long} + \bar{p}, \qquad p_{long} = -\frac{\partial \varphi_{long}}{\partial t}, \tag{55}$$

$$\bar{p} = -\frac{gk_0(1+\sigma^2)}{4\sigma}|A|^2\frac{\cosh(2k_0(z+h))}{\cosh(2k_0 h)} - \frac{\partial \bar{\varphi}}{\partial t}$$

$$- \frac{g}{4\sigma^2 V}\mathrm{Im}\left[A^*\frac{\partial A}{\partial t}\right]\left(\rho_1\frac{\sinh 2k_0(z+h)}{\cosh 2k_0 h} + \rho_2\frac{\cosh 2k_0(z+h)}{\cosh 2k_0 h} + \rho_3\right),$$

$$\rho_1 = 2\sigma(1+\sigma^2)k_0(z+h),$$

$$\rho_2 = -(1+\sigma^2)(k_0 h(1+\sigma^2) - \sigma),$$

$$\rho_3 = -(1-\sigma^2)(k_0 h(1+\sigma^2) - \sigma).$$

The solution of the system (48)-(50) and the corresponding contribution to the pressure in (55) may be calculated similar to the deep-water limit (34),

$$\frac{\partial \overline{\varphi}}{\partial t} = \left(\frac{\omega_0}{2\sigma} + \frac{k_0 V(1-\sigma^2)}{4\sigma}\right) \int_{-\infty}^{\infty}\int_{-\infty}^{\infty} \omega |A(\tau)|^2 \frac{\cosh\left(\frac{\omega}{V}(z+h)\right)}{\sinh\left(\frac{\omega}{V}h\right)} e^{i\omega(t-\tau)} d\omega d\tau . \qquad (56)$$

According to (53)-(55), the pressure at the bottom, $z = -h$, is given by the following expressions,

$$p^{(I)}(z=-h) = \sqrt{1-\sigma^2}\, g\, \mathrm{Re}(AE) - h\sigma\sqrt{1-\sigma^2}\, \frac{g}{V}\, \mathrm{Im}\left(\frac{\partial A}{\partial t} E\right) \qquad (57)$$

$$p^{(II)}(z=-h) = \frac{g k_0 (1-\sigma^2)(3-4\sigma^2)}{4\sigma^3}\, \mathrm{Re}(A^2 E^2), \qquad (58)$$

$$\overline{p}(z=-h) = -\frac{g k_0 (1-\sigma^2)}{4\sigma}|A|^2 - \left.\frac{\partial \overline{\varphi}}{\partial t}\right|_{z=-h} + \frac{(1-\sigma^2)(k_0 h(1+\sigma^2)-\sigma)}{\sigma^2}\, \frac{g}{2V}\, \mathrm{Im}\left[A^* \frac{\partial A}{\partial t}\right]. \qquad (59)$$

As discussed in the Appendix, the obtained solution formally corresponds to the second-order theory for the first and the second harmonic, and to the third-order theory for the low-frequency part. However, under the condition of deep water, the factor $(1-\sigma^2)$ becomes anomalously small. As a result, the second term in $\overline{p}$ (59) may become greater than the first one if $(1-\sigma^2) \sim 4\exp(-2k_0 h) < \varepsilon$; the third term may exceed the first one when $(k_0 h)^{-1} < \varepsilon$. The values of $k_p h$ and corresponding $(1-\sigma^2)$ are given in Table 1 for the simulated cases assuming $k_0 = k_p$, and $k_p$ is calculated from $\omega_p$ according to the linear dispersion relation (10). The steepness parameter $\varepsilon$ is also estimated in Table 1 as $k_p H/2$ for the regular waves A-E and via the root-mean-square elevation $k_p \eta_{rms}$. One may see that the ratio between $\varepsilon$ and $(1-\sigma^2)$ may be different in the simulated cases, thus all the terms in (55) will be taken into account in what follows. The second harmonic $p^{(II)}$ vanishes in the limit $\sigma \to 1$ ($k_0 h \to \infty$), its effect on the pressure beneath waves is expected to be less important, as it quickly attenuates with depth.

The effect of the wave modulation on the dynamic pressure enters the solution through the second-order correction to the first harmonic (53) and the third-order correction to the low-frequency part (55), it produces the mean flow $\overline{\varphi}$. Unlike the Dysthe model, in the finite depth theory the wave nonlinearity produces a correction to the mean surface displacement (43) and the pressure (see the first term in (59)) even when the wave is uniform. Then the term $\overline{\eta}$ corresponds to the difference between the mean water level and the rest water level.

**Second harmonic of the pressure**

The expression for the second harmonic $p^{(II)}$ (54) consists of two terms. The first one comes from the nonlinear surface boundary condition, it decays with depth with the scale $(2k_0)^{-1}$ according to the Laplace equation; the corresponding nonlinear correction in the Bernoulli equation is negligibly small. The second term in $p^{(II)}$ originates from the nonlinear part of the Bernoulli equation. It does not attenuate with depth, but its magnitude is proportional to $(1-\sigma^2)$, which vanishes in the limit of infinite depth, $\sigma \to 1$. Thus, under the condition of finite depth the nonlinear character of the Bernoulli law (6) cannot be ignored when analyzing the second-order nonlinear pressure field.

Let us consider a uniform wave with the amplitude $a$. The amplitude of its second harmonic we estimate with the help of the solution for $\eta^{(II)}$ (42). Then the linear theory for the pressure (9) will lead to the following solution for the second harmonic,

$$p^{(II)}_{linear} = \frac{gk_0(3-\sigma^2)}{4\sigma^3}\frac{\cosh(2k_0(z+h))}{\cosh(2k_0h)}a^2\cos(2\omega_0 t - 2k_0 x). \tag{60}$$

Meanwhile the nonlinear theory (54) yields the following expression for the second pressure harmonic,

$$p^{(II)}_{nonlinear} = \frac{gk_0(1-\sigma^2)}{4\sigma^3}\left[3(1+\sigma^2)\frac{\cosh(2k_0(z+h))}{\cosh(2k_0h)} - \sigma^2\right]a^2\cos(2\omega_0 t - 2k_0 x). \tag{61}$$

Depending on the choice of $h$ and $z$, $-h \leq z \leq 0$, the relation between the values of the linear (60) and nonlinear (61) solution may be quite different. When the nonlinear solution (61) is divided by the linear solution (60), the ratio at the bottom, $z = -h$, reads

$$\left.\frac{p^{(II)}_{nonlinear}}{p^{(II)}_{linear}}\right|_{z=-h} = \frac{(1+\sigma^2)(3-4\sigma^2)}{3-\sigma^2}, \tag{62}$$

where the identity

$$\cosh(2k_0 h) = \frac{1+\sigma^2}{1-\sigma^2} \tag{63}$$

has been used. The ratio (62) is equal to zero when $3 - 4\sigma^2 = 0$ ($k_0h \approx 1.32$); it is positive for shallower basins and negative for sufficiently deep waters, $k_0h > 1.32$. The ratio (62) is equal to 1 only in very shallow water (see the solid line in Fig. 4) and tends to $-1$ in the limit of infinite depth. Hence, the absolute values of the linear and nonlinear solutions coincide in the deep water limit, though the signs are opposite.

It may be straightforwardly shown that the sign of the vertical mode of the solution (61) is determined by the function

$$f(z) = \cosh^2(k_0(z+h)) - \frac{1}{6}\cosh^2(k_0 h) - \frac{1}{3}, \tag{64}$$

which is monotonically growing when $z$ increases from $z = -h$ to $z = 0$. Hence, the minimum of $f$ is given by $f(z = -h) = (4 - \cosh^2 k_0 h)$; it is also easy to see that $f(0) > 0$. Therefore, the function $f(z)$ may change the sign in the interval $[-h, 0]$ in sufficiently deep basins, when $\cosh k_0 h > 2$, i.e., $k_0h > 1.32$. The horizon where $p^{(II)}$ equals to zero is specified by the condition $f = 0$. This point is close to the bottom if $k_0h$ slightly exceeds the value of 1.32, and is located at the horizon $k_0 z \approx -0.5 \ln 6 \approx -0.90$ in the limit of infinite depth, $k_0 h \to \infty$. Thus, the second harmonic of the pressure beneath uniform waves should change its sign within the interval of dimensionless depths $0.90 < |k_0 z| < 1.32$ in general. The coefficient in square brackets of the expression for $p^{(II)}$ (54) is positive when close to the water surface.

Similar to (62), the ratio of the nonlinear and linear pressures calculated at the still water horizon $z = 0$ reads

$$\left.\frac{p^{(II)}_{nonlinear}}{p^{(II)}_{linear}}\right|_{z=0} = \frac{(1-\sigma^2)(3+2\sigma^2)}{3-\sigma^2}, \tag{65}$$

which is always positive but decays to zero when the depth is large (see the dashed line in Fig. 4). When the dimensionless water depth is $k_0h = 2$, the ratio (65) is less than 0.17; it drops down to about 0.02 at $k_0h = 3$ and further diminishes under deeper water conditions.

**Low-frequency part of the pressure**

As discussed above, the solution $\bar{p}$ (55), reproduced from (A.38), is accurate to the order $O(\varepsilon^3)$; it includes one term $O(\varepsilon^2)$ and two terms of formally smaller magnitude, $O(\varepsilon^3)$. The latter terms tend to the solution of the Dysthe theory (33) in the limit of a very large depth, $k_0 h \to \infty$. The term $\partial \bar{\varphi} / \partial t$ is determined by the problem (48)-(50), its closed form is given in (56).

Let us consider the case when the wave train is almost uniform so that the first term in $\bar{p}$ (55) knowingly dominates. If the long wave is characterized by a single wavenumber $K$, then $p^{(0)}$ (55) reads

$$p^{(0)} = g\eta_{long} \frac{\cosh(K(z+h))}{\cosh(Kh)} + g\bar{\eta} \cosh(2k_0(z+h)). \qquad (66)$$

Since $\bar{\eta}$ is always negative, the induced pressure $\bar{p}$ leads to generally smaller total pressure than compared to the hydrostatic problem of the pressure beneath the displacement $\bar{\eta}$. The nonlinear pressure $\bar{p}$ approaches hydrostatic with depth [Dean & Dalrymple, 2010]. In the limit of a very long wave, $Kh \to 0$, the first summand in (66) may be approximated by the hydrostatic solution $g\eta_{long}$, and then the bottom pressure produced by the long wave and by the nonlinear wave set-down cannot be distinguished,

$$p^{(0)}(z = -h) \approx g(\eta_{long} + \bar{\eta}) \quad \text{if} \quad Kh \to 0. \qquad (67)$$

In the deep-water limit $k_0 h \to \infty$, the set-down $\bar{\eta}$ vanishes, and the second summand in (66) tends to $-\omega_0^2/2|A|^2 \exp(2k_0 z)$ (the corresponding term is present in the solution (33)), which quickly attenuates with depth.

Let us estimate the difference between the linear and nonlinear solutions for the low frequency harmonic of the pressure beneath uniform waves with the amplitude $a$, similar to Eqs. (60) and (61). The zeroth harmonic (produced by a constant elevation) does not attenuate with depth. According to the linear solution,

$$p^{(0)}_{linear} = g\bar{\eta} = -\frac{gk_0(1-\sigma^2)}{4\sigma} a^2, \qquad (68)$$

where $\bar{\eta}$ is taken from (43). According to (55), the nonlinear solution reads

$$p^{(0)}_{nonlinear} = -\frac{gk_0(1+\sigma^2)}{4\sigma} a^2. \qquad (69)$$

Therefore their relation is constant along the depth,

$$\frac{p^{(0)}_{nonlinear}}{p^{(0)}_{linear}} = \frac{(1+\sigma^2)}{(1-\sigma^2)} = \cosh(2k_0 h), \qquad (70)$$

hence the magnitudes of the nonlinear and linear solutions for the zero-frequency pressure differ drastically when $k_0 h$ grows.

## 6. Comparison of the laboratory measurements with the new two-component theory

In this section, we compare the two-component finite-depth nonlinear theory (2CT) for the pressure beneath surface waves, derived in Sec. 5, with the experimental data, assuming $k_0 = k_p$. At first, we consider the experiments with regular waves, when the solution (41)-(46), (53)-(55) simplifies, putting $\partial A/\partial t = 0$:

$$\eta^{(I)} = \operatorname{Re}(AE), \quad \eta^{(II)} = \frac{k_0(3-\sigma^2)}{4\sigma^3} \operatorname{Re}(A^2 E^2), \qquad (71)$$

$$\eta^{(0)} = \eta_{long} + \bar{\eta}, \quad \bar{\eta} = -\frac{k_0(1-\sigma^2)}{4\sigma}|A|^2,$$

$$\varphi^{(I)} = -\frac{g}{\omega_0}\operatorname{Im}(AE)\frac{\cosh(k_0(z+h))}{\cosh(k_0 h)}, \quad \varphi^{(II)} = -\frac{3\omega_0(1-\sigma^4)}{8\sigma^4}\operatorname{Im}(A^2 E^2)\frac{\cosh(2k_0(z+h))}{\cosh(2k_0 h)}, \qquad (72)$$

$$\varphi^{(0)} = \varphi_{long}, \quad \bar{\varphi} = 0,$$

$$p^{(I)} = g \operatorname{Re}(AE) \frac{\cosh(k_0(z+h))}{\cosh(k_0 h)}, \tag{73}$$

$$p^{(II)} = \frac{gk_0(1-\sigma^2)}{4\sigma^3}\left[3(1+\sigma^2)\frac{\cosh(2k_0(z+h))}{\cosh(2k_0 h)} - \sigma^2\right]\operatorname{Re}(A^2 E^2),$$

$$p^{(0)} = p_{long} + g\overline{\eta}\cosh(2k_0(z+h)).$$

As discussed in Sec. 5, the long wave contributes linearly to the low frequency parts of the surface displacement $\eta^{(0)}$, velocity potential $\varphi^{(0)}$ and the pressure $p^{(0)}$. If $\eta_{long}$ and $\varphi_{long}$ are zero, the solution (71)-(73) is nothing else but the second-order solution for the Stokes wave in finite depth, which may be found in [Demirbilek & Vincent, 2002; Pinto & Neves, 2003].

At first, we check the amplitudes of the first three harmonics of the surface displacement, which we denote as $A^{(0)}$, $A^{(I)}$ and $A^{(II)}$, calculated from the experimental data. Since the wave train is assumed uniform, the amplitude of the zeroth harmonic is calculated as the mean value of the analyzed sample of the surface displacements, $A^{(0)}_{exp} = <\eta(t)>$. Here, the amplitudes of the first and second harmonics are calculated using the Fourier transform with application of the Hann filter. As the signal samples are not perfectly periodic, the spectral peaks in the Fourier transforms are smeared and possess slowly decaying tails. The use of the Hann smoothing allows more accurate determination of the harmonic amplitudes. The amplitudes estimated from the experimental records are given in Table 2 with subscripts "*exp*".

The theoretical amplitudes $A^{(0)}$, $A^{(I)}$, $A^{(II)}$ follow from the solution (71),

$$A^{(I)} = A_0, \quad A^{(II)} = \frac{k_p(3-\sigma^2)}{4\sigma^3}A_0^2, \quad A^{(0)} = \langle \eta_{long}\rangle + \overline{\eta}, \quad \overline{\eta} = -\frac{k_p(1-\sigma^2)}{4\sigma}A_0^2, \tag{74}$$

where the constant $A_0$ characterizes the regular wave amplitude. Assuming that the analyzed wave samples are sufficiently short, the long wave surface displacement is replaced by its mean value, $\eta_{long} = <\eta_{long}>$. The theoretical amplitudes $A^{(II)}_{theor}$ and $<\eta_{long}> = A^{(0)}_{exp} - \overline{\eta}$ are calculated according to (74) for the experimental values $A_0 = A^{(I)}_{exp}$. The results are listed in Table 2; the shadowed cells denote the cases when the measured signal was close to the level of noise. In the series A the second harmonic could not be discerned. In the series D and E the application of the Hann filter allows to discern up to the fourth nonlinear harmonic of the surface displacement. The magnitudes of experimentally measured second harmonics correspond to the theoretical values with the accuracy about 10% in the cases D and E; the agreement is noticeably worse in the cases B, C. The estimated values of $\overline{\eta}$ (74) were almost zeros, therefore the registered non-zero mean surface displacement is presumably produced by the parasitic long-scale motions of the water, $<\eta_{long}> \approx A^{(0)}_{exp}$.

The difference between the measured pressure time series and the solution (73) evaluated as the root mean square difference (12) is shown in Fig. 2 with blue triangles. Similar to the case when the Dysthe model was applied (Sec. 4), the complex amplitude $A(t)$ is obtained using an iterative procedure; its Fourier transform is confined within the spectral band $[\omega_p/2, 3\omega_p/2]$. The pressure $p_{long}$, produced by the estimated free long wave, is calculated using the linear solution (51) for the Fourier components within the frequency interval from 0 to $\omega_p/2$. One may conclude that the new theory exhibits generally the best agreement with the experimental data in the entire water column (except the case A with obscure results), particularly in the "difficult" cases of steep waves D-F.

A deeper analysis of the regular wave cases A-E is provided in Fig. 5, where profiles of the pressure vertical modes are plotted for the low-frequency part, and for the first and second harmonics. The full circles correspond to the experimental data; the thick red curves are the solution (73) where the long-wave term is calculated as $p_{long} = g<\eta_{long}>$. The dotted gray curves show the linear solution (9) calculated for the harmonic amplitudes $A^{(0)}_{exp}$, $A^{(I)}_{exp}$, $A^{(II)}_{exp}$ respectively. According to the linear solution, the zero-frequency mode does not depend on depth. As the surface displacement records, which correspond to the pressure measurements at different horizons, are not identical (as discussed in Sec 2), the curves of the linear solution exhibit some rough variation with depth (especially for the low-frequency part, see the left panels). The bars along the right sides of the graphs indicate the pressure reconstruction errors. They are computed according to (12), where the numerator is calculated for the corresponding frequency bands $[0, \omega_p/2)$, $[\omega_p/2, 3\omega_p/2)$ and $[3\omega_p/2, 5\omega_p/2)$, while the denominator is calculated for all frequencies.

The difference between the measurements and the theoretical curves for the dominant wave (central column in Fig. 5) almost cannot be seen by a naked eye. In different experiments at different horizons the error *Err* of the reconstruction of the first harmonic makes no more than 6% of the total pressure field.

The second harmonic of the dynamic pressure in the small-amplitude experiment A could not be evaluated accurately due to too small magnitudes. In the other small-amplitude case C at shallower condition the linear and nonlinear theories, and also the experimental data exhibit approximate correspondence. The experimental second harmonic in the deeper-water case B is significantly smaller than the linear solution (thin gray lines), but is a few times larger than the nonlinear solution.

In the experiments D and E ($k_p h \approx 1.7$, intense waves) the profile of the second harmonic calculated according to the new theory follows the experimental data remarkably well. The modulus of the theoretical amplitude of the second harmonic is plotted in the figures; hence, the theoretical curve touches the zero at about $k_p z \approx -0.9$ as discussed in Sec. 5. The experimental results seem to exhibit qualitatively similar dependencies, though the minimum amplitude of the second pressure harmonic is observed at somehow shallower horizons. At larger depths the vertical mode of the second harmonic remains approximately constant. The linear solution for the second harmonic is much larger in magnitude than the experimental data when close to the surface, though it decreases down to the values similar to the nonlinear solution, in agreement with the discussion in Sec. 5.

In general, the second harmonic contains little energy; as a result, its imperfect reconstruction contributes to the total error with no more than 2% in all the cases of regular waves A-E.

As follows from Fig. 5, the biggest reconstruction error is caused by the low-frequency part of the dynamic pressure. Focusing on the large-amplitude cases of regular waves (B, D, E), one may see that the theoretical modes generally agree with the observations, but the low-frequency component amplitude is underestimated by the new theory in the shallower cases D and E. Unlike the experimental data, the linear solution for the low-frequency part is just a constant value $g<\eta_{long}>$; it fluctuates in Fig. 5 from one horizon to another due to slightly different associated series of the surface displacements, as discussed earlier. Hence, the account of the freely propagating long wave is important; it noticeably reduces the error of reconstruction.

We have performed a further study in order to understand the nature of the drastic difference between the theories in describing modulated waves, observed in Fig. 2F. At first, we checked the frequency spectrum of the reconstructed surface displacement as shown in Fig. 6 for the representative cases E and F. The Fourier amplitudes of the experimental time series are given by the thick black curves. The reconstructed fields (from the fitted complex amplitudes $A(t)$) are shown with the pink areas for the Dysthe model (left) and for the new two-component theory for the modulated train (right). The numbers above the plots display the reconstruction errors in the corresponding frequency bands: low frequency, first and second harmonics, and also at higher frequencies. The new theory does not describe the third harmonic, but it is small in amplitude. Similar errors in the corresponding frequency bands are observed in the two nonlinear theories for the given experiment, E or F. According to Fig. 6, the reconstruction error grows significantly in the broad-band case F; the low-frequency part yields the most significant difference by 20%. It is readily seen in Fig. 6 that in every case a significant part of the low-frequency domain is not described by the theoretical fit (the filled areas are well below the black curves in the interval $\omega < \omega_0/2$). In the two-component theory presented in Sec. 5, this residual part of the low frequency surface displacement is associated with the long wave component $\eta_{long}$. The surface displacements, reconstructed in the form of truncated series, $\bar{\eta} + \eta^{(I)} + \eta^{(II)} + \eta^{(III)}$ for the Dysthe theory (red solid line) and $\eta_{long} + \bar{\eta} + \eta^{(I)} + \eta^{(II)}$ for the new theory (black dashed line), are compared in Fig. 3a with the original record of the steep wave event. Neither of the theoretical curves reproduces in full the sharp crests, but the low-frequency variations are noticeably better described by the two-component model. As a result, the pressure records in Fig. 3b,c are better reconstructed by the new theory.

Finally, in Fig. 7 we display the plots similar to Fig. 2F for the irregular wave case, but for specific frequency bands: the low frequency (left), the dominant band (middle), and the second harmonic (right). According to the plots, the description of the first harmonic by the new theory is better than by the Dysthe theory (a few percent, most likely due to the account of the finite depth in the coefficients). As for the second harmonic, according to Fig. 7 the difference between the nonlinear theories is small, while the linear theory serves a little bit better. A deeper investigation reveals that in this frequency domain freely propagating waves dominate over the nonlinear second harmonic due to the broad spectrum of the train. Therefore the linear solution is the most significant.

The low frequency part exhibits the biggest difference in Fig. 7: the Dysthe model describes the measurement much worse than the other theories; the corresponding error grows with depth. The new two-component nonlinear theory describes the low frequency part of the pressure field noticeably better, in particular in the upper water layer. We have checked that if the long-wave term is cancelled in the solution (55), $p_{long} = 0$, or if it is evaluated using the mean elevation, $p_{long} = g<\eta_{long}>$, then the low-frequency solution of the new theory almost coincides with the Dysthe theory. Hence the crucial improvement compared to the Dysthe model is achieved because the residual low-frequency surface displacement is assigned to the other, freely propagating long wave.

Convergence of the asymptotic series is examined in Fig. 8 by the example of the extreme event in the series F. The solution (53) for the first harmonic of the pressure is represented by terms of two asymptotic orders (the first and the second line of Eq. (53) respectively), $p^{(I)} = p_1^{(I)} + p_2^{(I)}$, $p_1^{(I)} \sim O(\varepsilon)$, $p_2^{(I)} \sim O(\varepsilon^2)$. These functions are plotted in Fig. 8 for two locations: close to the surface and at the deepest probe. One can clearly see that the discrimination between the magnitudes of the asymptotic orders weakens at the deeper location. The situation is in fact even worse within the Dysthe theory (30), which involves three asymptotic orders. Hence, the poor convergence of the asymptotic theories at great depths worsens the theoretical description. This circumstance can explain the uneven behavior of the functions of errors for the nonlinear theories at the deepest horizons in Fig. 7 (middle panel) and the growing with depth errors of the Dysthe theory in Fig. 2c-f.

## 7. Conclusions

In this work we have analyzed the experimental data of simultaneous recording of the surface water displacement and pressure fields beneath planar regular and irregular waves of different intensity. The experimental conditions correspond to intermediate water depths, when the Stokes wave representation should be applicable. The purpose was to check, how accurately the pressure field could be reconstructed having the surface displacement time series in hand (a single-point direct problem), using the asymptotic weakly nonlinear theory for slowly modulated waves. The linear theory for potential waves was used as a classical reference case.

The performed study reveals that the linear theory serves rather well for the description of the pressure beneath waves in deep and intermediate depth conditions even when the waves are steep. At the same time, the vertical dependence of the nonlinear pressure harmonics is described by the linear solutions incorrectly. While the error associated with the high frequency part of the nonlinear dynamic pressure is not large, the slowly attenuating with depth low-frequency pressure is the main source of discrepancy. It is shown that the use of the traditional deep-water nonlinear Dysthe model can improve the description of the dynamic pressure close to the surface, but yields unacceptably poor pressure reconstruction at great depths.

A new two-component theory for weakly modulated weakly nonlinear waves in water of finite depth is proposed (41)-(56), which reproduces the second-order solution from Brinch-Nielsen & Jonsson (1986), and takes into consideration an extra long wave component. The freely propagating long wave may be naturally excited in the experimental wave tank due to various uncontrolled effects. The two-component theory yields significantly better description of the low-frequency pressure.

According to the finite-depth second-order theory, the vertical mode of the second harmonic of the dynamic pressure crosses zero between $k_p h \approx 0.90$ and $k_p h \approx 1.32$. The direct measurements of narrow-banded nonlinear waves exhibit reasonable agreement with this prediction. The linear solution for the dynamic pressure yields unrealistically large amplitude of the second harmonic, what can be a source of large errors when solving the inverse problem using the linear transfer function.

The new theory exhibits an overall better description of the pressure field beneath nonlinear waves. In the considered examples the largest root-mean-square error of the pressure reconstruction by the new theory lies within 16%, including the sequence with the Yura wave. The corresponding samples of the reconstructed time series (Fig. 3) look pretty well.

**Acknowledgement**

The work was supported by the state contract No 0030−2021−0007 (AVS) and by the RFBR projects 19-55-15005 and 20-05-00162 (AVK).

**Appendix. Derivation of the weakly nonlinear theory for weakly modulated waves in water of finite depth**

In this section we give details of the derivation of the weakly nonlinear theory for the pressure beneath planar surface gravity waves, which propagate in a basin with relatively large constant depth. The governing system of equations is given by (1)-(4). The wave train will be assumed weakly nonlinear and slowly modulated, $k_0 a = O(\varepsilon)$, $k_0 L_{train} = O(\varepsilon^{-1})$, $\varepsilon \ll 1$. Here $k_0$ is the carrier wavenumber of the wave train, $a$ is its characteristic amplitude, $L_{train}$ is the characteristic length of the modulation. In the leading order the solution is sought in the following form, where fast $(x_0, t_0, z_0)$ and slow $(x_1, t_1, z_1)$ variables are distinguished,

$$\eta_{train}(x,t) = \varepsilon \, \text{Re}[A(x_1,t_1)E], \quad (A.1)$$

$$\varphi_{train}(x,t,z) = \varepsilon \, \text{Re}[B(x_1,t_1)\Phi_{10}(z_0)E], \quad (A.2)$$

$$E(x_0,t_0) \equiv \exp(i\omega_0 t_0 - ik_0 x_0), \quad (A.3)$$

Multiple-scale coordinates and times are introduced following the traditional asymptotic approach,

$$\frac{\partial}{\partial x} = \frac{\partial}{\partial x_0} + \varepsilon \frac{\partial}{\partial x_1}, \quad \frac{\partial}{\partial z} = \frac{\partial}{\partial z_0} + \varepsilon \frac{\partial}{\partial z_1}, \quad \frac{\partial}{\partial t} = \frac{\partial}{\partial t_0} + \varepsilon \frac{\partial}{\partial t_1}. \quad (A.4)$$

The wave train is characterized by the complex amplitude $A(x_1, t_1)$ which depends on slow coordinate and time, while the fast function $E(x_0, t_0)$ oscillates with the carrier wavenumber $k_0$ and the frequency $\omega_0$. The function $\Phi_{10}(z_0)$,

$$\Phi_{10}(z_0) = \frac{\cosh(k_0(z_0 + h))}{\cosh(k_0 h)}, \quad (A.5)$$

specifies the leading order vertical structure of the wave train. Only three harmonics will be taken into account:

$$\eta = \bar{\eta} + \eta^{(I)} + \eta^{(II)}, \quad \varphi = \bar{\varphi} + \varphi^{(I)} + \varphi^{(II)}. \quad (A.6)$$

The leading order solution (A.1), (A.2) corresponds to the first harmonic, $E^1$; the terms of the second harmonic are proportional to $E^2$ and $E^{-2}$; the terms with overheads correspond to the "difference" nonlinear harmonic $E^0$.

The following ansatz for the second-order solution is used, in line with the traditional derivation of the nonlinear Schrödinger equation and the Dysthe equation, e.g. [Slunyaev, 2005; Trulsen, 2006]

$$\eta^{(I)}(x,t) = \varepsilon \operatorname{Re}[A(x_1,t_1)E] + \varepsilon^2 \operatorname{Re}[A_{11}(x_1,t_1)E] + O(\varepsilon^3), \quad (A.7)$$

$$\varphi^{(I)}(x,t,z) = \varepsilon \operatorname{Re}[B(x_1,t_1)\Phi_{10}(z_0)E]$$
$$+ \varepsilon^2 \operatorname{Re}\left[B_{11}(x_1,t_1)\Phi_{10}(z_0)E + i\frac{\partial B}{\partial x_1}\Phi_{11}(z_0)E\right] + O(\varepsilon^3) \quad (A.8)$$

$$\eta^{(II)}(x,t) = \varepsilon^2 \operatorname{Re}[A_{21}(x_1,t_1)E^2] + O(\varepsilon^3), \quad (A.9)$$

$$\varphi^{(II)}(x,t,z) = \varepsilon^2 \operatorname{Re}[B_{21}(x_1,t_1)\Phi_{21}(z_0)E^2] + O(\varepsilon^3), \quad (A.10)$$

$$\bar{\eta}(x,t) = \varepsilon^2 A_{01}(x_1,t_1) + O(\varepsilon^3), \quad (A.11)$$

$$\bar{\varphi}(x,t,z) = \varepsilon^2 \varphi_{01}(x_1,t_1,z_1) + O(\varepsilon^3). \quad (A.12)$$

Our primary goal is to obtain the first contribution to the nonlinear second harmonic and to the difference harmonic in the finite-depth consideration.

We substitute the series (A.6)-(A.12) to the governing equations (1)-(4), where the boundary conditions at $z = \eta$ are approximated in the vicinity of the water rest level $z = 0$ using the Taylor expansions. After that, the terms of different powers of $\varepsilon$ and $E$ are collected and considered separately. The Laplace equation dictates the following form of the high-order corrections to the vertical mode structure:

$$\Phi_{21}(z_0) = \frac{\cosh(2k_0(z_0 + h))}{\cosh(2k_0 h)}, \quad \Phi_{11}(z_0) = (z_0 + h)\frac{\sinh(k_0(z_0 + h))}{\cosh(k_0 h)}. \quad (A.13)$$

Using the truncated series (A.7)-(A.12), non-zero components of the Laplace equation with harmonics $E^1$ and $E^2$ appear in the order $O(\varepsilon^3)$. The slow vertical dependence of the long-scale velocity potential component $\varphi_{01}$ (A.12) reflects the long horizontal scale of the induced current.

For the zeroth harmonic the lowest order of the Laplace equation yields

$$\varepsilon^4 \left[\frac{\partial^2 \varphi_{01}}{\partial x_1^2} + \frac{\partial^2 \varphi_{01}}{\partial z_1^2}\right] = 0, \quad -h \leq z \leq 0 \quad (A.14)$$

with the zero vertical gradient at the bottom,

$$\varepsilon^3 \frac{\partial \varphi_{01}}{\partial z_1} = 0, \qquad z = -h. \tag{A.15}$$

In the order $O(\varepsilon)$ the first harmonic of the surface boundary conditions gives the relation between the surface displacement and the velocity potential for the wave train,

$$B(x_1, t_1) = i \frac{g}{\omega_0} A(x_1, t_1). \tag{A.16}$$

The dispersion relation which coincides with (10) follows as well,

$$\omega_0^2 = g k_0 \sigma, \qquad \sigma \equiv \tanh k_0 h. \tag{A.17}$$

In the order $O(\varepsilon^2)$ the first harmonic of the dynamic boundary condition yields the relation

$$A_{11} = -i \frac{\omega_0}{g} B_{11} + i h \sigma \frac{\partial A}{\partial x_1} - \frac{i}{\omega_0} \frac{\partial A}{\partial t_1}. \tag{A.18}$$

We require that the correction to the surface displacement $A_{11}$ is zero; then the quantity $B_{11}$ is determined,

$$B_{11} = \frac{g h \sigma}{\omega_0} \frac{\partial A}{\partial x_1} - \frac{g}{\omega_0^2} \frac{\partial A}{\partial t_1}. \tag{A.19}$$

Note that a different condition was implied in the solution by Brinch-Nielsen & Jonsson (1986). There the first terms from the right hand side of (A.18) were balanced. As a result, our theory reproduces the formulas by Brinch-Nielsen & Jonsson (1986) after an asymptotic transformation of the complex wave amplitude (see e.g. in Trulsen, 2006). The surface conditions in the order $O(\varepsilon^2)$ give the following relation for the difference harmonic:

$$A_{01}(x_1, t_1) = -\frac{k_0(1-\sigma^2)}{4\sigma} |A(x_1, t_1)|^2. \tag{A.20}$$

The dynamic condition in the order $\varepsilon^2 E^2$ yields

$$A_{21} = -2i \frac{\omega_0}{g} B_{21} - \frac{3 k_0 (1-\sigma^2)}{4\sigma} A^2. \tag{A.21}$$

The kinematic boundary condition in the order $\varepsilon^2 E^1$ relates the evolutions of $A(x,t)$ in time and space,

$$\frac{\partial A}{\partial t_1} + V \frac{\partial A}{\partial x_1} = 0, \qquad V = \frac{g}{2\omega_0}\left(\sigma + k_0 h (1-\sigma^2)\right). \tag{A.22}$$

In the advection equation (A.22) the term $V$ is equal to the wave group velocity, $V = d\omega/dk$, of the carrier wave with the wavenumber $k_0$.

In the order $\varepsilon^2 E^2$ the kinematic boundary condition yields

$$B_{21} = \frac{3\omega_0(1-\sigma^4)}{8\sigma^4} A^2. \tag{A.23}$$

The condition which determines the induced mean flow, appears in the order $\varepsilon^3 E^2$ of the kinematic boundary condition,

$$\left.\frac{\partial \varphi_{01}}{\partial z_1}\right|_{z=0} = \frac{g k_0}{2\omega_0} \frac{\partial |A|^2}{\partial x_1} - \frac{k_0(1-\sigma^2)}{4\sigma} \frac{\partial |A|^2}{\partial t_1}. \tag{A.24}$$

Combining the listed formulas, one can write down the solution (A.6) as follows,

$$\eta^{(I)} = \mathrm{Re}(AE) + O(\varepsilon^3), \tag{A.25}$$

$$\eta^{(II)} = \frac{k_0(3-\sigma^2)}{4\sigma^3} \mathrm{Re}(A^2 E^2) + O(\varepsilon^3), \tag{A.26}$$

$$\bar{\eta} = -\frac{k_0(1-\sigma^2)}{4\sigma} |A|^2 + O(\varepsilon^3), \tag{A.27}$$

$$\varphi^{(I)} = -\frac{g}{\omega_0}\operatorname{Im}(AE)\frac{\cosh(k_0(z+h))}{\cosh(k_0 h)} \tag{A.28}$$

$$-\frac{\omega_0}{2\sigma^2 k_0^2 V}\operatorname{Re}\!\left(\frac{\partial A}{\partial t}E\right)\!\left[(\sigma+k_0 h(1+\sigma^2))\frac{\cosh(k_0(z+h))}{\cosh(k_0 h)} - 2\sigma k_0(h+z_0)\frac{\sinh(k_0(z+h))}{\cosh(k_0 h)}\right] + O(\varepsilon^3),$$

$$\varphi^{(II)} = -\frac{3\omega_0(1-\sigma^4)}{8\sigma^4}\operatorname{Im}(A^2 E^2)\frac{\cosh(2k_0(z+h))}{\cosh(2k_0 h)} + O(\varepsilon^3), \tag{A.29}$$

The problem on the induced flow $\bar{\varphi}$ consists of the following set of equations,

$$\frac{1}{V^2}\frac{\partial^2 \bar{\varphi}}{\partial t^2} + \frac{\partial^2 \bar{\varphi}}{\partial z^2} = O(\varepsilon^5), \qquad z \le 0, \tag{A.30}$$

$$\frac{\partial \bar{\varphi}}{\partial z} = -\left(\frac{\omega_0}{2\sigma V} + \frac{k_0(1-\sigma^2)}{4\sigma}\right)\frac{\partial}{\partial t}|A|^2 + O(\varepsilon^4), \qquad z = 0, \tag{A.31}$$

$$\frac{\partial \bar{\varphi}}{\partial z} = 0, \qquad z = -h. \tag{A.32}$$

The physical coordinates and time are used in the expressions (A.25)-(A.32) and below. The relations (A.28), (A.30), (A.31) are formulated for the time domain using (A.22), bearing in mind the application to the laboratory data represented by time series.

The dynamic pressure (6) beneath the surface waves reads

$$p = \bar{p} + p^{(I)} + p^{(II)}, \tag{A.33}$$

$$p^{(I)} = g\operatorname{Re}(AE)\frac{\cosh(k_0(z+h))}{\cosh(k_0 h)} \tag{A.34}$$

$$+ \frac{g}{V}\operatorname{Im}\!\left(\frac{\partial A}{\partial t}E\right)\!\left[\frac{(z+h)\sinh(k_0(z+h))}{\cosh(k_0 h)} - \frac{h\sigma\cosh(k_0(z+h))}{\cosh(k_0 h)}\right] + O(\varepsilon^3),$$

$$p^{(II)} = \frac{gk_0(1-\sigma^2)}{4\sigma}\left[\frac{3(1+\sigma^2)}{\sigma^2}\frac{\cosh(2k_0(z+h))}{\cosh(2k_0 h)} - 1\right]\operatorname{Re}(A^2 E^2) + O(\varepsilon^3), \tag{A.35}$$

$$\bar{p} = -\frac{gk_0(1+\sigma^2)}{4\sigma}\frac{\cosh(2k_0(z+h))}{\cosh(2k_0 h)}|A|^2 + O(\varepsilon^3). \tag{A.36}$$

The solution for the pressure component $p^{(II)}$ (A.35) contains the factor $(1-\sigma^2)$, which quickly vanishes when the depth $k_0 h$ grows. The low-frequency part of the pressure $\bar{p}$ at the bottom gives

$$\bar{p}(z=-h) = -\frac{gk_0(1-\sigma^2)}{4\sigma}|A|^2 + O(\varepsilon^3), \tag{A.37}$$

with the same factor $(1-\sigma^2)$. As a result, these components become anomalously small in deep water; the terms of a formally higher order $O(\varepsilon^3)$ can prevail if $(1-\sigma^2) < \varepsilon$. The third-order terms may be particularly important for the slowly attenuating with depth low-frequency component produced by the difference between wave harmonics.

The next-order solution for $\bar{p}$ may be written based on the already calculated asymptotic series as follows,

$$\bar{p} = -\frac{gk_0(1+\sigma^2)}{4\sigma}|A|^2\frac{\cosh(2k_0(z+h))}{\cosh(2k_0 z)} - \frac{\partial \bar{\varphi}}{\partial t} \tag{A.38}$$

$$-\frac{g}{4\sigma^2 V}\operatorname{Im}\!\left[A^*\frac{\partial A}{\partial t}\right]\!\left(\rho_1\frac{\sinh 2k_0(z+h)}{\cosh 2k_0 h} + \rho_2\frac{\cosh 2k_0(z+h)}{\cosh 2k_0 h} + \rho_3\right) + O(\varepsilon^4),$$

$$\rho_1 = 2\sigma(1+\sigma^2)k_0(z+h),$$
$$\rho_2 = -(1+\sigma^2)(k_0 h(1+\sigma^2)-\sigma),$$
$$\rho_3 = -(1-\sigma^2)(k_0 h(1+\sigma^2)-\sigma).$$

In the limit $k_0 h \to \infty$ the solution for $\bar{p}$ (A.38) tends to the Dysthe's solution (33). A similar correction $O(\varepsilon^3)$ to the low-frequency part of $\bar{\eta}$ may be also obtained without derivation of higher order asymptotic expansions,

$$\bar{\eta} = -\frac{k_0(1-\sigma^2)}{4\sigma}|A|^2 - \frac{1}{g}\frac{\partial \bar{\varphi}}{\partial t}\bigg|_{z=0} \tag{A.39}$$
$$+ \frac{(1-\sigma^2)(k_0 h(1+\sigma^2)-\sigma)}{4\sigma^2 V}\operatorname{Im}\left[A^* \frac{\partial A}{\partial t}\right] + O(\varepsilon^4).$$

The solution for the surface displacement and pressure (A.25), (A.26), (A.34), (A.35), (A.38), (A.39) is used in this work; it represents the second-order theory for the dominant and second harmonics, and the third-order theory for the low-frequency part.

**References**

1. Adcock, T.A.A., Taylor, P.H. Non-linear evolution of uni-directional focussed wave-groups on a deep water: A comparison of models. Appl. Ocean Res 59, 147–152 (2016).

2. Ardhuin, F., Balanche, A., Stutzmann, E., Obrebski, M. From seismic noise to ocean wave parameters: General methods and validation. J. Geophys. Res. 117, C05002 (2012).

3. Bonneton, P., Lannes, D. Recovering water wave elevation from pressure measurements. J. Fluid Mech. 833, 399-429 (2017).

4. Brinch-Nielsen, U., Jonsson, I.G. Fourth order evolution equations and stability analysis for Stokes waves on arbitrary water depth. Wave Motion 8, 455-472 (1986).

5. Cavaleri, L., Ewing, J.A., Smith, N.D. Measurement of the Pressure and Velocity Field Below Surface Waves. In: Favre A., Hasselmann K. (eds) Turbulent Fluxes Through the Sea Surface, Wave Dynamics, and Prediction. Springer, Boston, MA (1978).

6. Clamond, D., New exact relations for easy recovery of steady wave profiles from bottom pressure measurements. J. Fluid Mech. 726, 547-558 (2013).

7. Clamond, D., Constantin A., Recovery of steady periodic wave profiles from pressure measurements at the bed. J. Fluid Mech. 714, 463-475 (2013).

8. Clamond, D. and Henry, D., Extreme water-wave profile recovery from pressure measurements at the seabed. J. Fluid Mech. 903, R3 (2020).

9. Clauss, G.F., Stempinski, F., Klein, M. Experimental and numerical analysis of steep wave groups. Proc. 12$^{th}$ Int. Congress Int. Maritime Association of the Mediterranean. (Varna, Bulgaria, 2-6 September, 2007).

10. Dean, R.G., Dalrymple, R.A., Water wave mechanics for engineers and scientists. 2010.

11. Demirbilek, Z., Vincent, L., Water Wave Mechanics. Coastal Engineering Manual Outline, S. Army Corps of Engineers, Washington, DC, Chapter. 11-1, 1-35 (2002).

12. DiMarco, S.F., Meza, E., Zhang, J. Estimating wave elevation from pressure using second order nonlinear wave-wave interaction theory with applications to hurricane Andrew. J. Coastal Res. 17, 658-671 (2001).

13. Dysthe, K.B. Note on a modification to the nonlinear Schrödinger equation for application to deep water waves. Proc. Roy. Soc. London A 369, 105-114 (1979).

14. Glubokovskikh, S., Pevzner, R., Sidenko, E., Tertyshnikov, K., Gurevich, B., Shatalin, S., Slunyaev, A., Pelinovsky, E. Downhole distributed acoustic sensing reveals the wave eld structure of the coastal microseisms. Journal of Geophysical Research - Solid Earth (2021, Submitted).

15. Fenton, J.D., A fifth-order Stokes theory for steady waves. J. Warterw. Port Coast. Ocean Eng., ASCE 111, 216–234 (1985).

16. Hasselmann, K. Ocean wave sources of seismic noise. Rev. Geophys. 1, 177–210 (1963).

17. Henry, D., Thomas, G.P., Prediction of the free-surface elevation for rotational water waves using the recovery of pressure at the bed. Phil. Trans. R. Soc. A 376, 20170102 (2018).

18. Kovalev, D.P., Dolgikh, G.I., Shevchenko G.V. Generation of low frequency microseisms by infragravity waves on the Southeastern coast of Sakhalin Island. Doklady Earth Sciences 461, 368–371 (2015).

19. Landès, M., Hubans, F., Shapiro, N.M., Pau, A., Campillo, M. Origin of deep ocean microseisms by using teleseismic body waves. J. Geophys. Res. 115, B05302 (2010).

20. Mori, N., Liu, P., Yasuda, T. Analysis of freak wave measurements in the Sea of Japan. Ocean Eng. 29, 1399–1414 (2002).

21. Mouragues, A., Bonneton, P., Lannes, D., Castelle, B., Marieu, V. Field data-based evaluation of methods for recovering surface wave elevation from pressure measurements. Coastal Eng. 150, 147-159 (2019).

22. Obrebski, M.J., Ardhuin, F., Stutzmann, E., Schimmel, M. How moderate sea states can generate loud seismic noise in the deep ocean. Geophys. Res. Lett. 39, L11601 (2012).

23. Pellet, L., Christodoulides, P., Donne, S., Bean, C.J., Dias, F. Pressure induced by the interaction of water waves with nearly equal frequencies and nearly opposite directions. Theor. Appl. Mech. Lett. 7, 138-144 (2017).

24. Pinto, F.T., Neves, A.C.V. Second order analysis of dynamic pressure profiles, using measured horizontal wave flow velocity component. In: Proc. 6th Int. Conf. on Computer Modelling Experimental Measurements of Seas and Coastal Regions, Environmental Studies Series Vol. 9, edited by Brebbia, C. A., Almorza, D., and Lopez-Aguayo, F. WIT Press, 237–252 (2003).

25. Shemer, L, Sergeeva, A., Slunyaev, A. Applicability of envelope model equations for simulation of narrow-spectrum unidirectional random field evolution: experimental validation. Phys. Fluids 22, 016601 (2010).

26. Slunyaev, A., Pelinovsky, E., Guedes Soares, C. Reconstruction of extreme events through numerical simulations. J. Offshore Mechanics and Arctic Engineering 136, 011302 (2014).

27. Slunyaev, A., Pelinovsky, E., Hsu, H.-C. The pressure field beneath intense surface water wave groups. Eur. J. Mech. B / Fluids 67, 25-34 (2018).

28. Slunyaev, A.V. A high-order nonlinear envelope equation for gravity waves in finite-depth water. JETP 101, 926-941 (2005).

29. Tang, T., Li, Y., Bingham H.B., and Adcock T.A.A. Comparison of two versions of the MNLS with full water wave equations. OMAE 2020 June 28-July 3 Fort Lauderdale, FL, USA (2020).

30. Touboul, J., Pelinovsky, E. On the use of linear theory for measuring surface waves using bottom pressure distribution. Eur. J. Mech. B Fluids 67, 97-103 (2018).

31. Trulsen, K. Weakly nonlinear and stochastic properties of ocean wave fields: application to an extreme wave event. In: Waves in geophysical fluids: Tsunamis, Rogue waves, Internal waves and Internal tides (Eds. Grue, J., Trulsen, K.), CISM Courses and Lectures No. 489. New York, Springer Wein. 2006.

32. Trulsen, K., Dysthe, K.B. A modified nonlinear Schrödinger equation for broader bandwidth gravity waves on deep water. Wave Motion 24, 281–289 (1996).

33. Tsai, C.H., Huang, M.C., Young, F.J., Lin, Y.C., Li, H.W. On the recovery of surface wave by pressure transfer function. Ocean Eng. 32, 1247–1259 (2005).

34. Vasan, V., Oliveras, K., Henderson, D., Deconinck, B. A method to recover water-wave profiles from pressure measurements. Wave Motion 75, 25-35 (2017).

35. Xiao, W., Liu, Y., Wu, G., Yue, D.K.P. Rogue wave occurrence and dynamics by direct simulations of nonlinear wave-field evolution. J. Fluid Mech. 720, 357-392 (2013).

36. Ying, Y., Bean, C.J., Bromirski, P.D. Propagation of microseisms from the deep ocean to land. Geophys. Res. Lett. 41, 6374–6379 (2014).

**Table 1.** Parameters of the measured wave sequences.

| Series | A | B | C | D | E | F |
|---|---|---|---|---|---|---|
| $h$, m | 0.40 | 0.40 | 0.40 | 0.40 | 0.40 | 0.40 |
| $T_p$, s | 0.67 | 0.67 | 1.00 | 1.00 | 1.00 | 0.73 |
| $k_p$, rad/m | 9.07 | 9.07 | 4.30 | 4.28 | 4.28 | 7.55 |
| $H/2$, m | 0.0015 | 0.015 | 0.0038 | 0.028 | 0.037 | |
| $\eta_{rms}$, m | 0.0011 | 0.010 | 0.0027 | 0.020 | 0.027 | 0.012 |
| $k_p h$ | 3.63 | 3.63 | 1.72 | 1.71 | 1.71 | 3.02 |
| $1-\sigma^2$ | 0.003 | 0.003 | 0.12 | 0.12 | 0.12 | 0.01 |
| $k_p H/2$ | 0.014 | 0.14 | 0.016 | 0.12 | 0.16 | |
| $k_p \eta_{rms}$ | 0.010 | 0.093 | 0.012 | 0.087 | 0.12 | 0.09 |

**Table 2.** Amplitudes of harmonics of the surface displacements. The gray shading indicates the cases when the level of signal was close to the noise.

| Series | $A^{(I)}_{exp}$, m | $A^{(II)}_{exp}$, m | $A^{(II)}_{theor}$, m | $A^{(0)}_{exp}$, m | $<\eta_{long}>$, m |
|---|---|---|---|---|---|
| A | $1.5 \cdot 10^{-3}$ | — | $1.1 \cdot 10^{-5}$ | $2.7 \cdot 10^{-5}$ | $2.7 \cdot 10^{-5}$ |
| B | $1.5 \cdot 10^{-2}$ | $2.1 \cdot 10^{-4}$ | $9.6 \cdot 10^{-4}$ | $-9.0 \cdot 10^{-5}$ | $-8.8 \cdot 10^{-5}$ |
| C | $3.8 \cdot 10^{-3}$ | $1.2 \cdot 10^{-5}$ | $4.1 \cdot 10^{-5}$ | $-1.4 \cdot 10^{-5}$ | $-1.2 \cdot 10^{-5}$ |
| D | $2.9 \cdot 10^{-2}$ | $1.9 \cdot 10^{-3}$ | $2.3 \cdot 10^{-3}$ | $-4.4 \cdot 10^{-4}$ | $-3.2 \cdot 10^{-4}$ |
| E | $3.8 \cdot 10^{-2}$ | $3.7 \cdot 10^{-3}$ | $3.9 \cdot 10^{-3}$ | $-51.2 \cdot 10^{-4}$ | $-3.2 \cdot 10^{-4}$ |

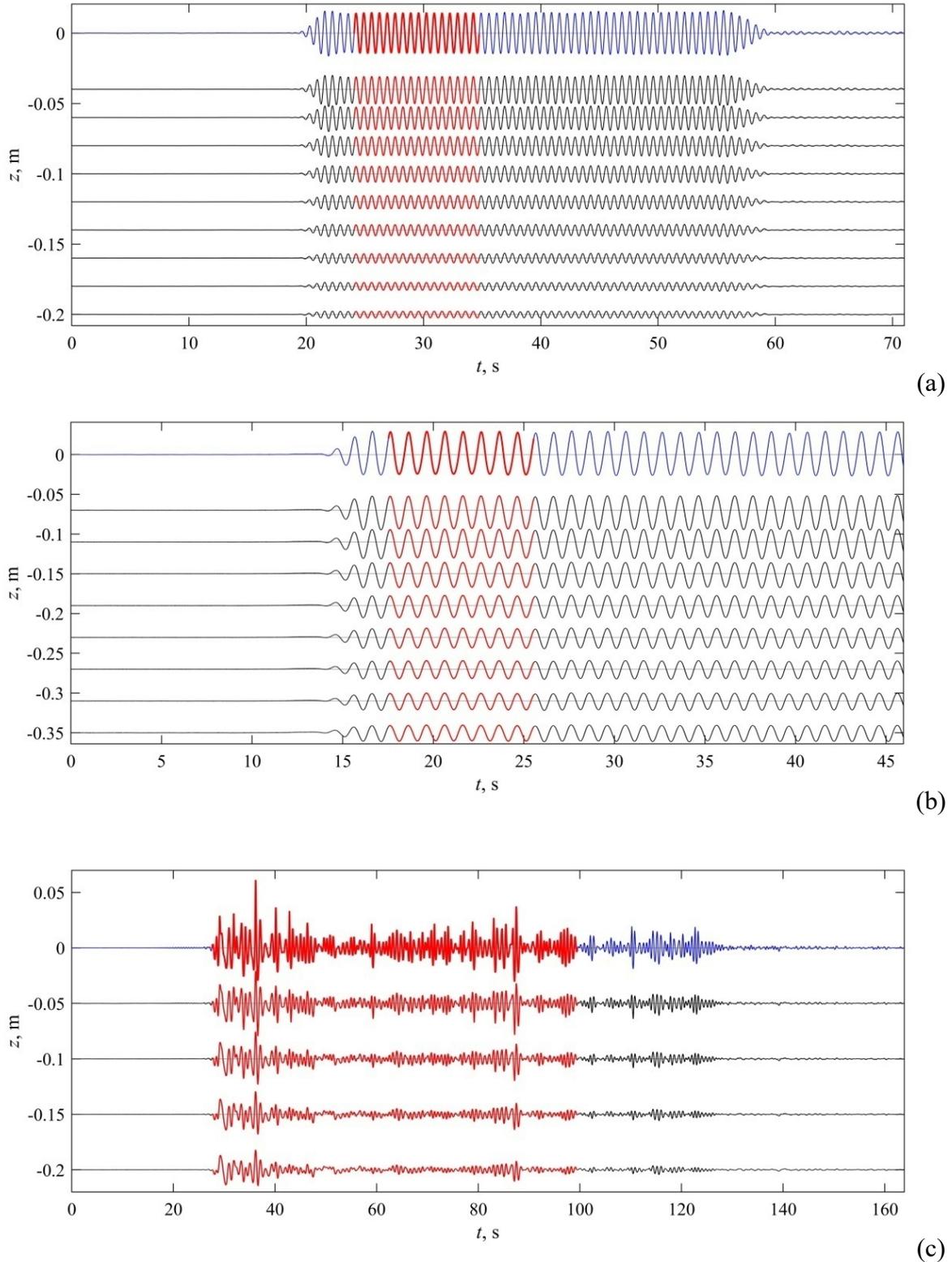

**Figure 1.** Time series of the surface displacement (upper curve) and the pressure records at a few horizons $z = -d$ (lower curves) measured in the laboratory experiments B (a), E (b) and F (c). The intervals extracted for the analysis are shown with the red color. Amplitudes of the time series are normalized for better visualization.

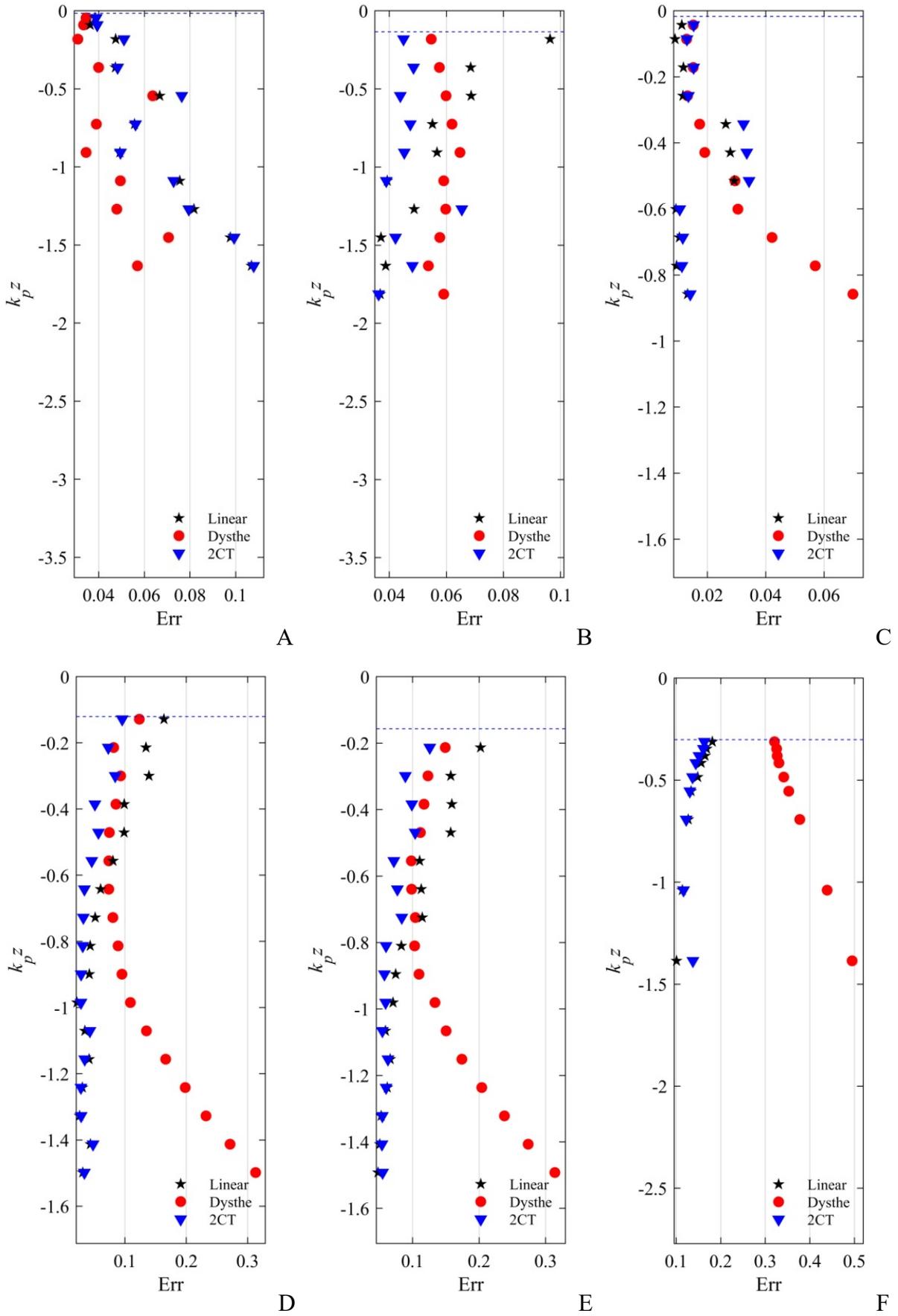

**Figure 2.** The differences *Err* (12) between the measured dynamic pressure and the pressures reconstructed from the time series of the surface displacement according to different theories. The horizontal dashed lines correspond to the horizons of deepest wave troughs. The minimum limits of the plots correspond to the bottoms.

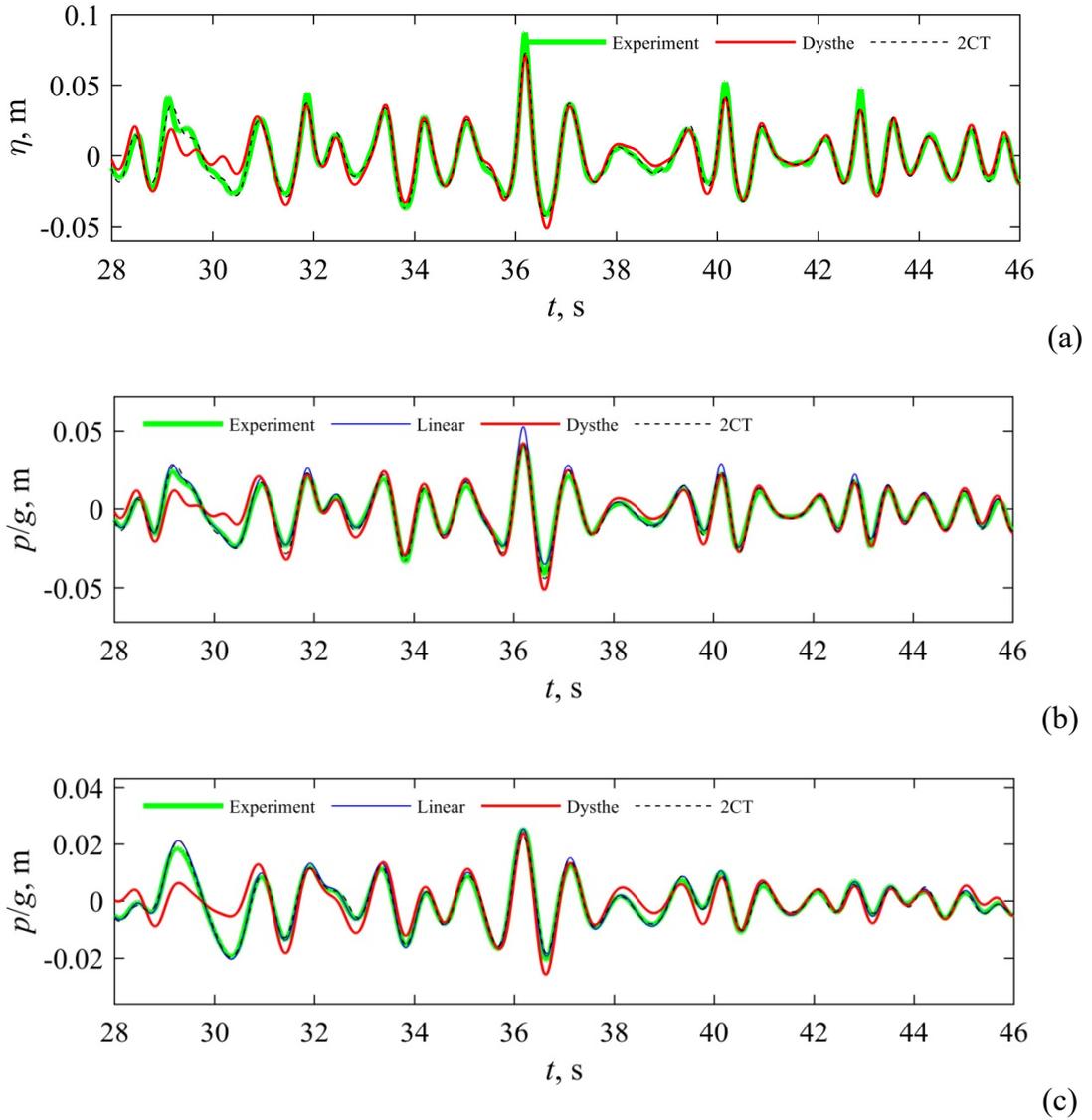

**Figure 3.** Performance of the theories to describe the interval of an extreme event in the series F: the measured and reconstructed surface displacement (a) and the dynamic pressures at the horizons $k_p z = -0.38$ (b) and $k_p z = -1.13$ (c).

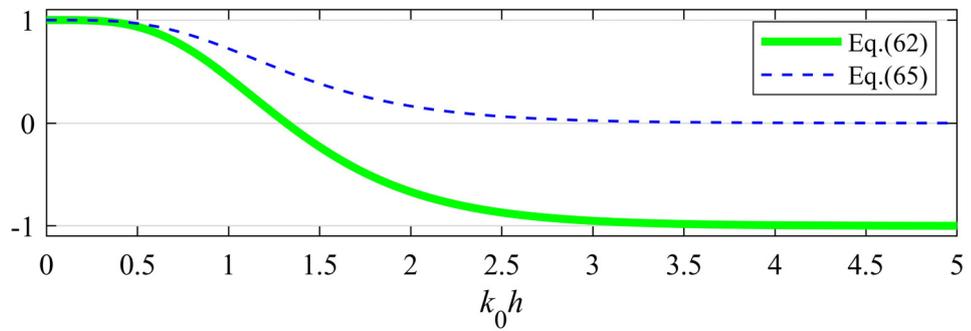

**Figure 4.** The ratios (62) and (65) as functions of the scaled depth $k_0 h$.

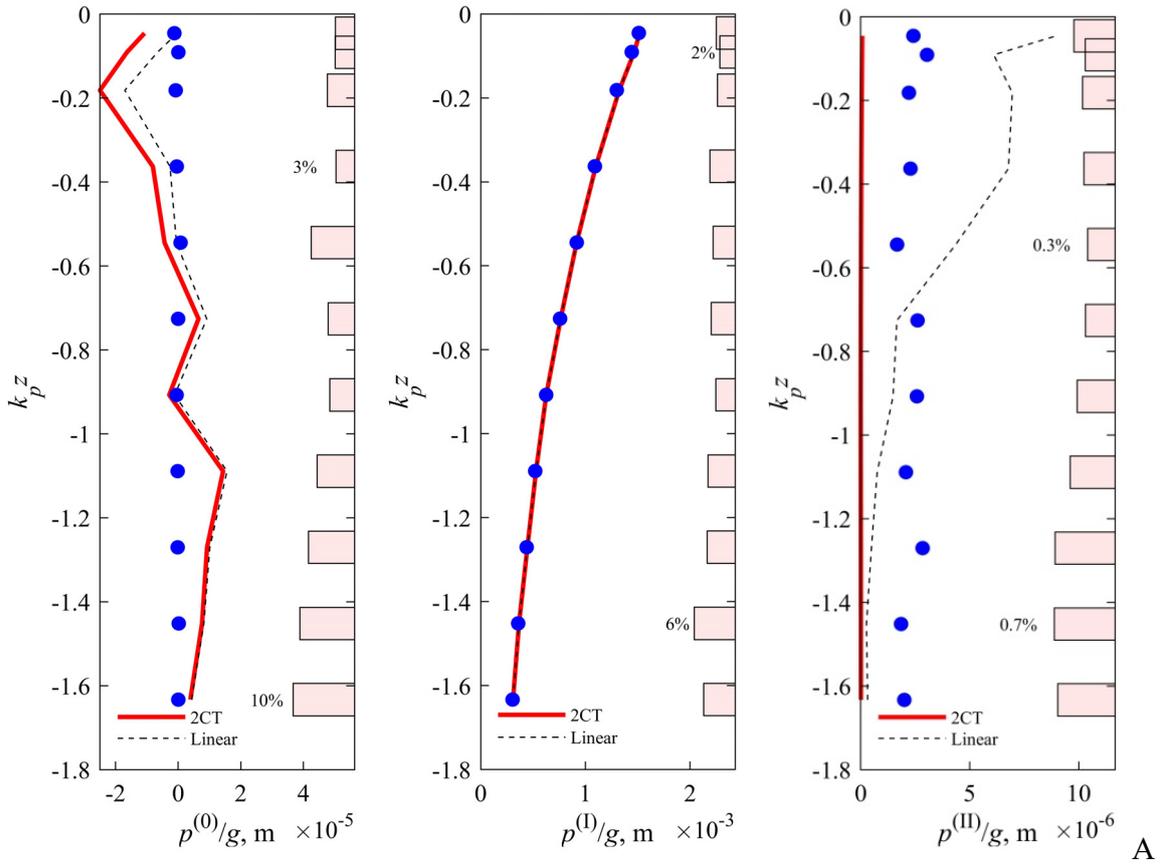
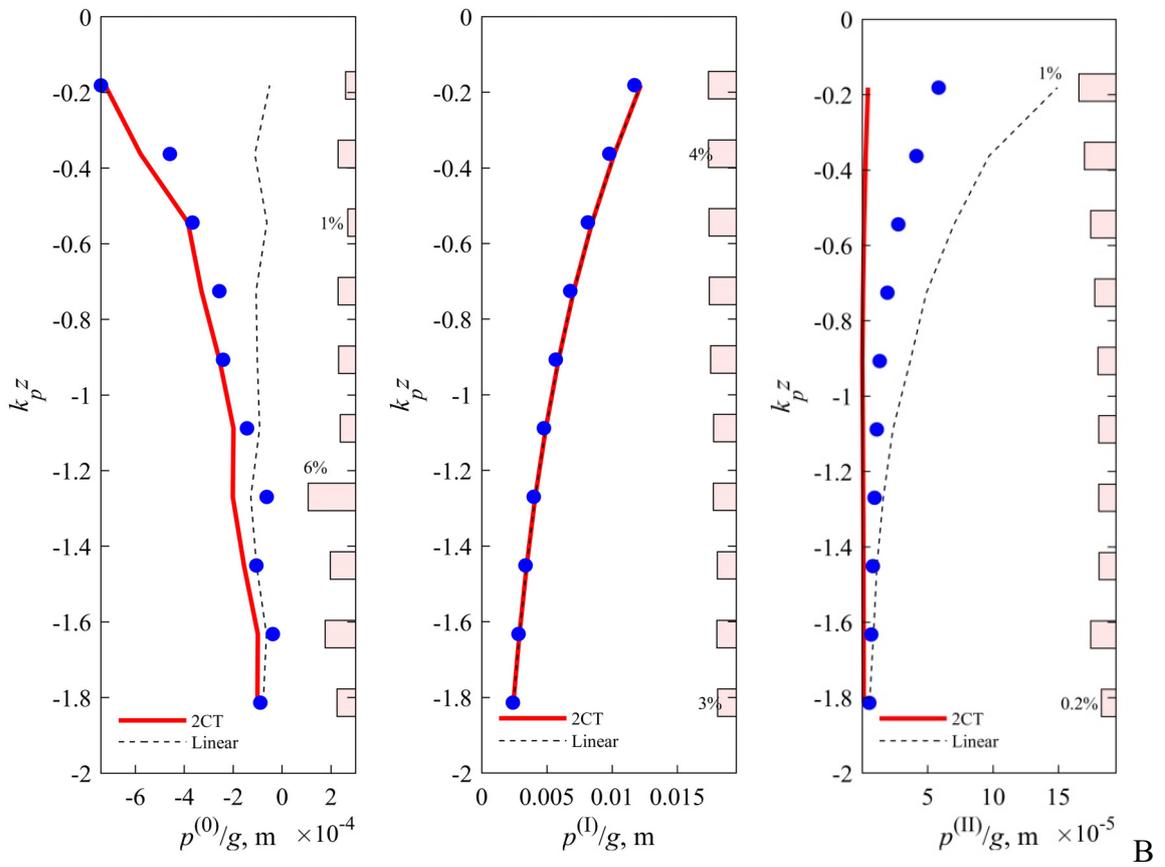

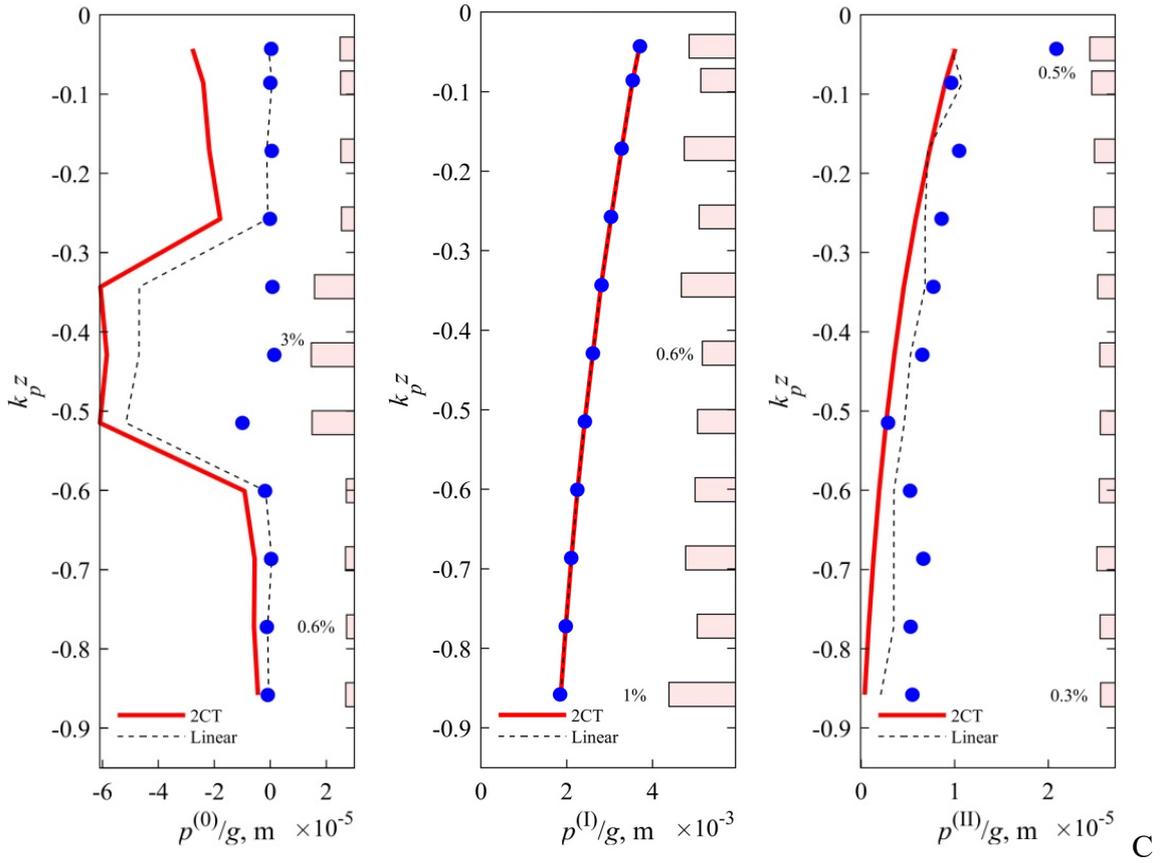

C

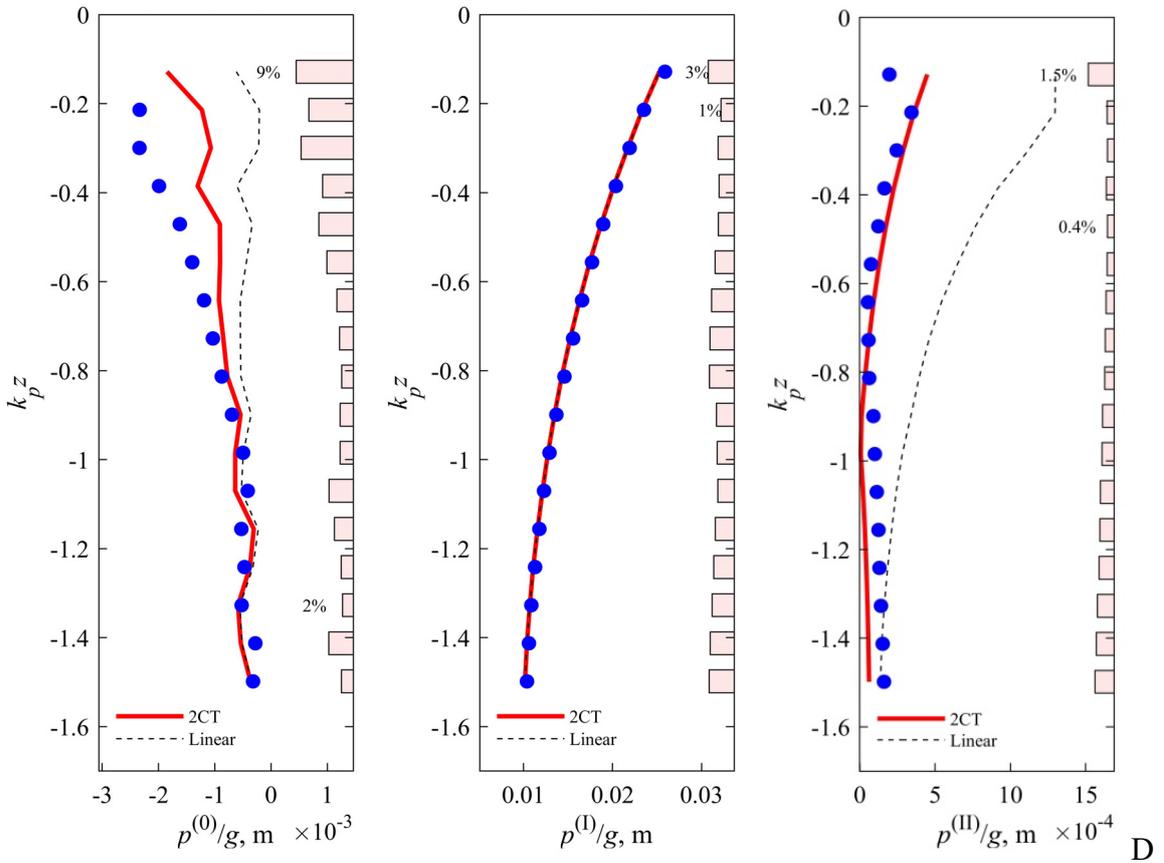

D

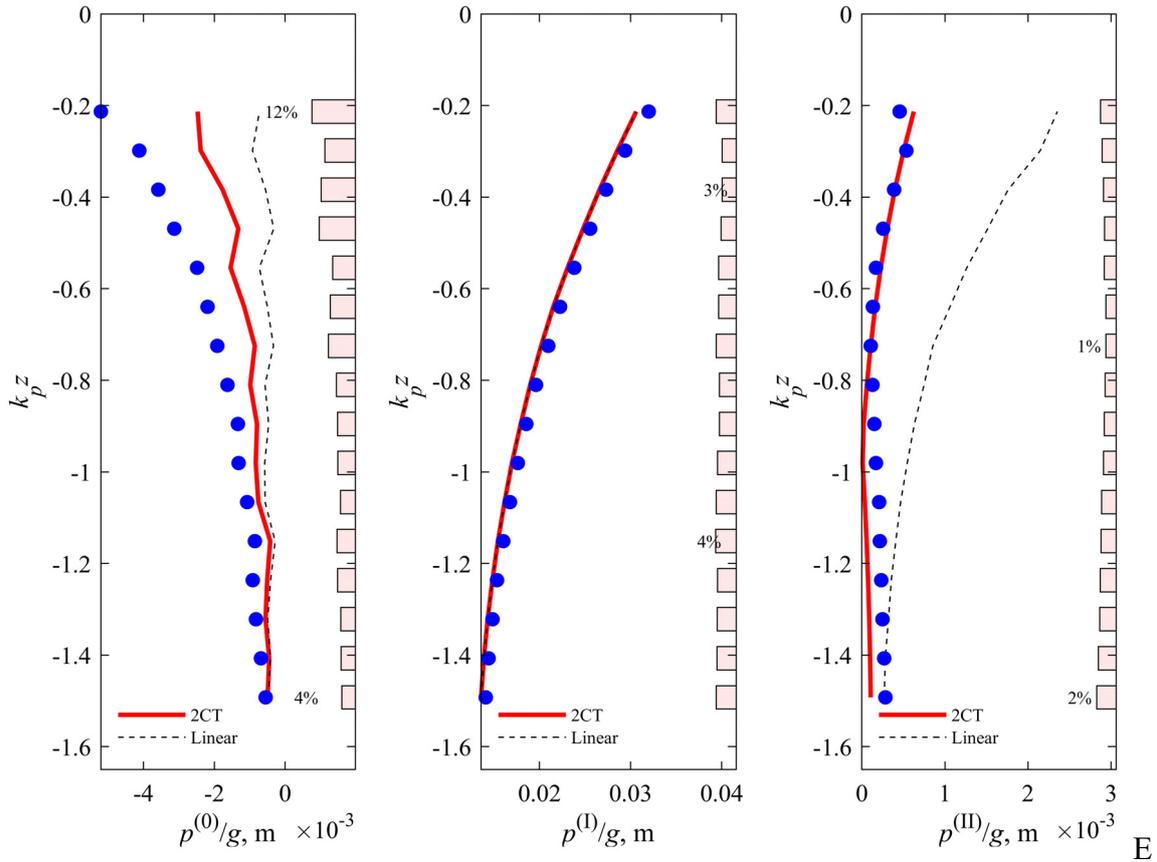

**Figure 5.** Mode structures of the low-frequency (left), dominant (center) and second (right) harmonics according to the instrumental measurements (full circles), the new two-component finite-depth theory (thick red line) and the linear solution (thin gray line). The bars to the right show the corresponding errors *Err* of the pressure reconstruction.

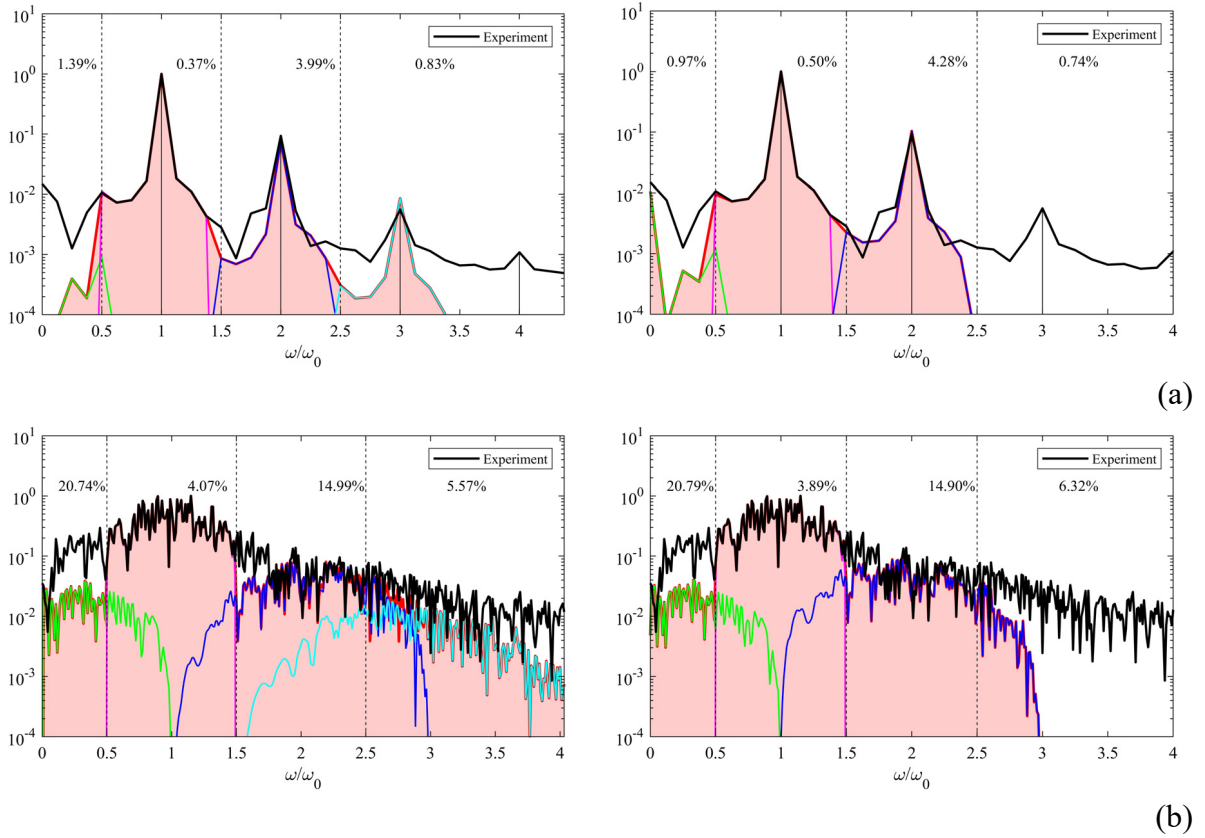

**Figure 6.** Fourier amplitudes of the measured surface displacement (thick black line) and of the fitted theoretical surface $\eta_{train}$ (shaded area) for the series E (a) and F (b). The left column corresponds to the Dysthe theory, while the right column to the new finite-depth theory. By different colors the curves for $\bar{\eta}$, $\eta^{(I)}$, $\eta^{(II)}$, $\eta^{(III)}$ are plotted (green, pink, blue and cyan respectively). The numbers on the top indicate the errors in the corresponding frequency bands: $(0, \omega_0/2)$, $(\omega_0/2, 3\omega_0/2)$, $(3\omega_0/2, 5\omega_0/2)$ and $\omega > 5\omega_0/2$.

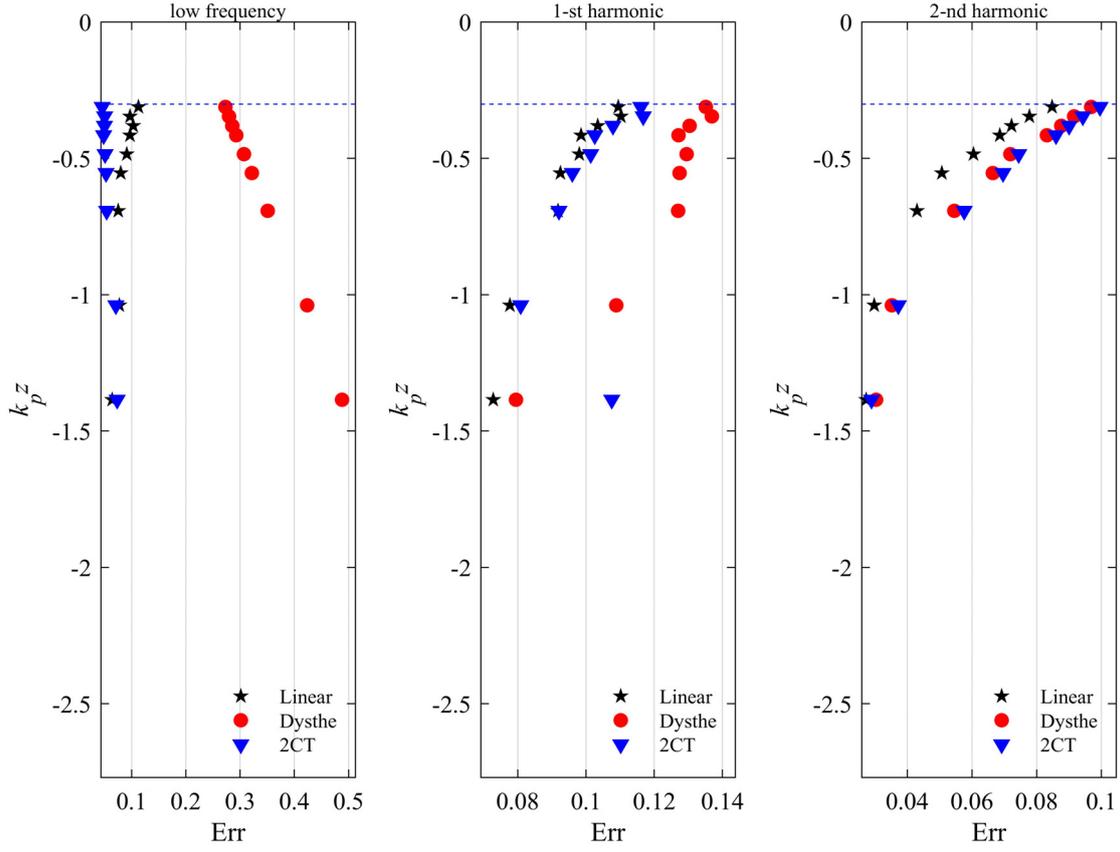

**Figure 7.** Same as Fig. 2 for the series F, but calculated for specific frequency bands $(0, \omega_p/2)$, $(\omega_p/2, 3\omega_p/2)$ and $(3\omega_p/2, 5\omega_p/2)$ from left to right correspondingly.

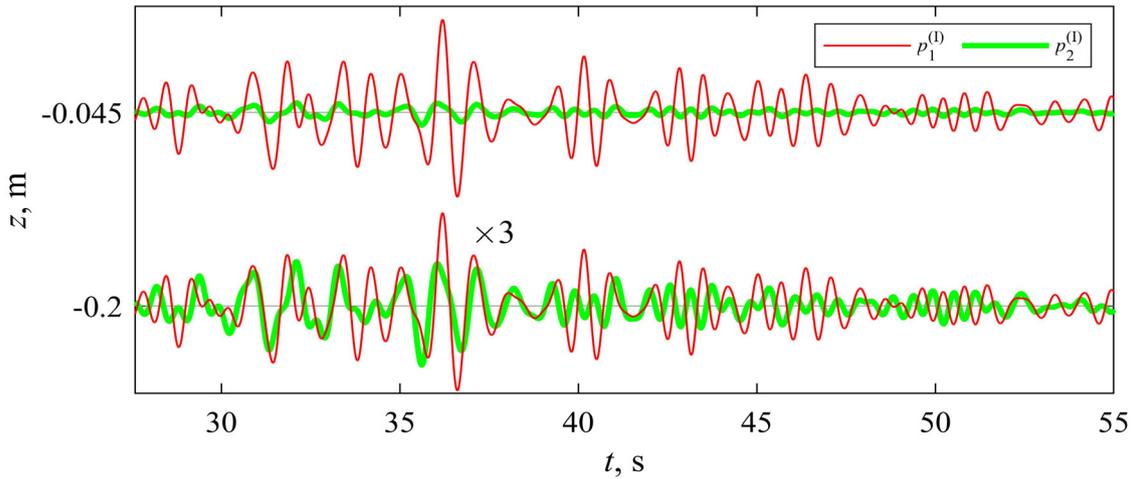

**Figure 8.** Components of two asymptotic orders for the first harmonic of the pressure (53) $p^{(I)} = p_1^{(I)} + p_2^{(I)}$ near the surface (above) and at the deepest probe (below, thrice amplified) for the extreme event in the series F.

33